\documentclass[10pt,journal,comsoc]{IEEEtran}

\usepackage[T1]{fontenc}

\usepackage{amsmath}
\usepackage[cmintegrals]{newtxmath}

\usepackage[english]{babel}
\usepackage{cite}
\usepackage[shortcuts, acronym, toc]{glossaries}
\newacronym
 {MIMO}{MIMO}{Multiple-Input Multiple-Output}

\newacronym
{SISO}{SISO}{Single-Input Single-Output}

\newacronym
{SIMO}{SIMO}{Single-Input Multiple-Output}

\newacronym
{MISO}{MISO}{Multiple-Input Single-Output}

\newacronym{MF}{MF}{Matched Filter}

\newacronym{WF}{WF}{Wiener Filter}

\newacronym
{MMSE}{MMSE}{Minimum Mean Squared Error}

\newacronym
{MSE}{MSE}{Minimum Squared Error}

\newacronym
{MSM}{MSM} {Maximum Safety Margin}

\newacronym
{SU}{SU}{Single-User}

\newacronym
{MU}{MU}{Multi-User}

\newacronym
{PSK}{PSK}{Phase-Shift Keying}

\newacronym
{QAM}{QAM}{Quadrature Amplitude Modulation}

\newacronym
{mmW}{mmW}{millimeter-Wave}

\newacronym
{ADC}{ADC}{Analog-to-Digital Converter}

\newacronym
{PA}{PA}{Power Amplifier}

\newacronym
{DAC}{DAC}{Digital-to-Analog Converter}

\newacronym
{MAC}{MAC}{Multiple Access Channel}

\newacronym
{BC}{BC}{Broadcast Channel}

\newacronym
{SR}{SR}{Symbol Region}

\newacronym
{BS}{BS}{Base Station}

\newacronym
{SNR}{SNR}{Signal-to-Noise Ratio}

\newacronym
{CE}{CE}{Constant Envelope}

\newacronym
{QCE}{QCE}{Quantized Constant Envelope}

\newacronym
{BER}{BER}{Bit Error Ratio}

\newacronym
{wrt}{w.r.t.}{with respect to}

\newacronym{CV}{CV}{Coefficient of Variation}
\newacronym{3D}{3D}{3 Dimensional}

\newacronym{AWGN}{AWGN}{Additive White Gaussian Noise}

\newacronym{iid}{i.i.d.}{independent and identically distributed}

\newacronym{R1}{R1}{rank one}

\newacronym{FR}{FR}{full rank}

\newacronym{PF}{PF}{Primal Feasibility}

\newacronym{DF}{DF}{Dual Feasibility}

\newacronym{CS}{CS}{Complementary Slackness}

\newacronym{KKT}{KKT}{Karush-Kuhn-Tucker}

\newacronym{ES}{ES}{Exhaustive Search}

\newacronym{OPIP}{OP/IP}{Optimisation Problem using Interior Point}

\newacronym{MUI}{MUI}{Multi-User Interference}

\newacronym{QN}{QN}{Quantisation Noise}

\newacronym{LB}{LB}{Lower Bound}

\newacronym{SQINR}{SQINR}{Signal-to-Quantisation-plus-Interference-plus-Noise Ratio}

\newacronym{SLNR}{SLNR}{Signal-to-Leakage-plus-Noise Ratio}

\newacronym{wsr}{wsr}{Weighted Sum-Rate}

\newacronym{WL}{WL}{widely-linear}

\newacronym{5G}{5G}{5$ ^{\text{th}} $ Generation Wireless Systems}

\newacronym{RF}{RF}{Radio Frequency}

\newacronym{CSI}{CSI}{Channel State Information}

\newacronym{PAPR} {PAPR}{Peak-to-Average-Power Ratio}

\newlength\maxlength
\newlength\thislength

\newglossarystyle{glossaryTableStyle}
{%
	{
		\setlength{\maxlength}{0pt}%
		\forglsentries[\currentglossary]{\thislabel}%
		{%
			\settowidth{\thislength}{\glsentryshort{\thislabel}}%
			\ifdim\thislength>\maxlength
			\setlength{\maxlength}{\thislength}%
			\fi
		}%
		\settowidth{\thislength}{\hspace{1.5em}=\hspace{1em}}%
		\setlength{\glsdescwidth}{\linewidth-\maxlength-\thislength-2\tabcolsep}%
		\begin{tabular}{l@{\hspace{1.5em}\hspace{1em}}p{\glsdescwidth}}
			\toprule
			\multicolumn{1}{l}{\textbf{Acronym}} &
			\multicolumn{1}{@{}l}{\textbf{Definition}}\\%
			\midrule
		}%
		{
			\bottomrule
		\end{tabular}%
	}%
	\renewcommand*{\glsgroupheading}[1]{}%
}

\usepackage{tikz,tkz-euclide}
\usetikzlibrary{arrows,shapes,positioning,arrows,decorations.markings,calc}
\usetikzlibrary{dsp,chains}
\usepackage{color}
\definecolor{green}{rgb}{0,0.4,0.05}
\definecolor{red}{rgb}{0.8,0,0}

\usepackage{slashbox}
\usepackage{pgfplots}
\usepackage{authblk}
\usepackage{amsbsy}

\usepackage{fixmath}
\usepackage{amssymb}
\usetikzlibrary{spy}
\usepackage{lipsum}
\usepackage{mathtools}

\usepackage{cuted}

\usepackage{psfrag}
\usepackage{algorithm}
\usepackage{algpseudocode}
\usepackage{scalefnt}
\usepackage{defs}

\begin{document}
\title{Quantized Constant Envelope Precoding with PSK and QAM Signaling}

\author[1]{Hela~Jedda}
\author[2]{Amine~Mezghani}
\author[3]{A.~Lee~Swindlehurst}
\author[1,4]{Josef~A.~Nossek}
\affil[1]{Associate Professorship of Signal Processing, Technical University of Munich, 80290 Munich, Germany}
\affil[2]{Wireless Networking and Communications Group, University of Texas at Austin, Austin, TX 78712, USA}
\affil[3]{Center for Pervasive Communications and Computing, University of California, Irvine, Irvine, CA 92697, USA}
\affil[4]{Department of Teleinformatics Engineering, Federal University of Cear\'a, Fortaleza, Brazil}
\affil[ ] {Email: \{hela.jedda, josef.a.nossek\}@tum.de, amine.mezghani@utexas.edu, swindle@uci.edu}

\maketitle

\begin{abstract}
Coarsely quantized massive \gls{MIMO} systems are gaining more interest due to their power efficiency. We present a new precoding technique to mitigate the \gls{MUI} and the quantization distortions in a downlink \gls{MU} \gls{MIMO} system with coarsely \gls{QCE} signals at the transmitter. The transmit signal vector is optimized for every desired received vector taking into account the \gls{QCE} constraint. The optimization is based on maximizing the safety margin to the decision thresholds of the receiver constellation modulation. Simulation results show a significant gain in terms of the uncoded \gls{BER} compared to the existing linear precoding techniques.
\end{abstract}

\IEEEpeerreviewmaketitle
\glsresetall
\section{Introduction}
\label{sec:intro}
\IEEEPARstart{T}{he} next generation of mobile communication aims at increasing 1000-fold the network capacity, 10-100-fold the number of connected devices and decreasing 5-fold the latency time and the power consumption compared to 4G networks \cite{Osseiran.2014}. To achieve these challenging requirements, the following technologies are the subject of current research:
\begin{itemize}
\item massive \gls{MIMO} systems, where the base stations (BSs) are equipped with a very large number of antennas (100 or more) that can simultaneously serve many users \cite{Marzetta.2010,Rusek.2013,Hoydis.2013,Ngo.2013,Lu.2014},
\item \gls{mmW} communication, i.e. frequencies ranging between 30 GHz and 300 GHz, where the spectrum is less crowded and greater bandwidth is available \cite{Swindlehurst.2014, Niu.2015,Rappaport.2015} and
\item smaller cells with ranges on the order of 10-200 m, i.e. pico- and femtocells.
\end{itemize} 

First, massive \gls{MIMO} systems lead to a drastic increase in the number of \gls{RF} chains at the BS and hence in the number of the wireless front-end hardware components. Second, \gls{mmW} communication implies that the wireless front-end hardware components are operated at much higher frequencies. Third, reducing the cell size means that the number of cells per unit area is increased and thus results in a much more dense wireless network. Combining the three 
technologies means a dramatic increase in the number of \gls{RF} hardware 
elements operating at very high frequencies per unit area. Hence, the \gls{RF} power consumption per unit area alarmingly increases.  
While the above technologies are foreseen as key technologies for future communication systems, the increase in power consumption represents a crucial concern. 

The term green communication refers to the idea of minimizing power consumption while guaranteeing a certain quality of service. 
The most critical front-end elements in terms of power consumption, depending on whether the large number of antennas is situated at the transmitter or at the receiver, are the \glspl{ADC} in the uplink scenario, and mainly the \glspl{PA} and secondarily the \glspl{DAC} in the downlink scenario, which is the focus of this contribution. The PA is considered as the most power hungry device at the transmitter side \cite{Earth.2012,Blume.2010}. When the PA is run in the saturation region, i.e. the highly non-linear region, high power efficiency is achieved and hence less power is consumed \cite{Varahram.2014}. However, in the saturation region strong non-linear distortions are introduced to the signals. To avoid the PA distortions when run in the saturation region, we resort to PA input signals of \gls{CE}. CE signals have the property of constant magnitude leading to a unit \gls{PAPR}. Thus, the information is carried by the signal phase. 

To this end, polar (phase-based) \glspl{DAC} at the transmitter are designed to convert the time-discrete and value-discrete base-band signals into time-continuous but value-discrete, i.e. phase-discrete, \gls{CE} signals. The number of possible discrete phases is determined by the resolution of the \glspl{DAC}. The larger the resolution is, the more accurate the phase information at the \glspl{DAC}' outputs, but the larger their power consumption \cite{Cui.2005}.
To further reduce the hardware power consumption, the \glspl{DAC}' resolution can be reduced. The use of coarsely quantized \glspl{DAC} is also beneficial in terms of reduced cost and circuit area and can further simplify the surrounding \gls{RF} circuitry due to the relaxed linearity constraint, leading to very efficient hardware implementations. In this way, the power consumption is reduced twofold: power efficient \glspl{PA} due to the \gls{CE} signals and less power consuming \glspl{DAC} due to the low resolution. However, this approach leads to non-linear distortions that degrade the system performance and have to be mitigated by the precoder design in massive \gls{MU} \gls{MIMO} downlink systems.

\subsection{Related Works}
The first precoding techniques in the context of \gls{CE} transmit signals were introduced in \cite{Love.2003,Zheng.2007,Yu.2007}. The idea of \gls{CE} transmit signals was further exploited for massive MIMO systems in \cite{Mohammed.2012,Mohammed.2013,Mohammed.2013b, Mollen.2015}, where the \gls{MUI} is minimized subject to the \gls{CE} constraint, whereas recent works in \cite{Liu.2017} and \cite{Amadori.2017} exploit the constructive part of the \gls{MUI} to design the \gls{CE} precoder. The authors in \cite{Shen.2016} design a \gls{CE} precoder to maximize the \gls{SLNR}. In the above contributions, the \glspl{DAC} are assumed to have infinite resolution.

The contribution in \cite{Mezghani.2009} is an early work that addressed the precoding task with low resolution \glspl{DAC} at the transmitter. A linear \gls{MMSE} precoder is designed, while quantization distortion is taken into account. This precoding design is not in the context of coarsely \gls{QCE} signals since the \glspl{DAC} are not polar but cartesian (inphase- and quadrature- based). 
However, the extreme case of 1-bit \glspl{DAC} in \cite{Mezghani.2009} represents a special case of coarsely \gls{QCE} signals. Many contributions in the literature have studied this special case. They can be categorized in two groups: linear and non-linear precoders. In addition to the linear precoder in \cite{Mezghani.2009}, we introduced in \cite{Usman.2016} another linear precoder, where the second-order statistics of the 1-bit DAC signals are computed based on Price's theorem \cite{Price.1958}. Non-linear precoding techniques in this context were introduced in \cite{Jedda.2016, Jacobsson.2016, Jacobsson.2017, Jedda.2017, Castaneda.2017, Swindlehurst.2017}. The non-linear methods can be classified with respect to two design criteria: symbol-wise \gls{MSE} and symbol-wise \gls{MSM} exploiting the idea of constructive interference. In the context of the symbol-wise \gls{MSE}, the authors in \cite{Jacobsson.2016} presented a convex formulation of the problem and applied it to higher-order modulation scheme in \cite{Jacobsson.2017}. The problem formulation is based on semidefinite relaxation and squared $\ell_{\infty}$-norm relaxation. The same optimization problem was solved more efficiently in \cite{Castaneda.2017} and \cite{Castaneda.2017b}. 

In the context of symbol-wise \gls{MSM} in \cite{Jedda.2016}, we presented a precoding technique based on a minimum-bit-error-ratio criterion and made use of the box norm ($\ell_{\infty}$) to relax the 1-bit constraint. Recently, the work in \cite{Swindlehurst.2017} proposed a method to significantly improve linear precoding solutions in conjunction with 1-bit quantization by properly perturbing the linearly precoded signal for each given input signal to favorably impact
the probability of correct detection. In \cite{Jedda.2017} the safety margin to the decision thresholds of the received \gls{PSK} symbols is maximized subject to a relaxed 1-bit constraint using linear programming. The same optimization problem was solved by the Branch-and Bound algorithm in \cite{Landau.2017} for the particular QPSK case. To the best of our knowledge, the only works that considered the case of coarsely \gls{QCE} transmit signals are \cite{Noll.2017}, \cite{Jacobsson.2017c} and \cite{Nedelcu.2017}. In \cite{Noll.2017}, we propose a symbol-wise \gls{MSE} precoders based on the gradient-descent method under the strict \gls{CE} constraint or a relaxed polygon constraint. In \cite{Jacobsson.2017c}, the authors extend the method in \cite{Jacobsson.2016} to fit in the context of \gls{QCE} transmit signals. In \cite{Nedelcu.2017}, the authors use the greedy approach for the precoder design while having the symbol-wise \gls{MSE} as the design criterion, too. The contribution in \cite{Kazemi.2017} addresses the task of \gls{QCE} precoding in the context of using a single common \gls{PA} and separate digital phase shifters for the antenna front-ends. The optimization problem consists of designing the \gls{QCE} precoder while minimizing the \gls{MUI}, and the idea of constructive interference, \cite{Masouros.2013,Masouros.2015}, is not exploited as in our work. The concept of \gls{QCE} precoding and general constellations is studied in this contribution. It is worth mentioning that the \gls{QCE} precoding can be combined with appropriate pulse shaping strategies as in \cite{Jedda.2015,Jedda.2015b} to ensure an efficient spectral confinement. In \cite{Mollen.2016}, it was shown that \gls{CE} precoding is still power efficient even when considering the time processing. The same investigation can be conducted for the case of \gls{QCE} precoding. Here, we focus rather on the spatial design problem.

\subsection{Main Contributions} 
The main contributions in this paper are summarized as follows
\begin{enumerate}
	\item We propose a \gls{QCE} precoding in the context of massive MIMO systems, where the transmit signals have constant magnitude and phases are drawn from a discrete set. The precoder design exploits the idea of constructive interference and adapts the design criterion in \cite{Jedda.2017} to the coarsely \gls{QCE} case. The optimization problem is solved using linear programming, which is very advantageous in terms of complexity as it is one of the most widely applied and studied optimization techniques.
	\item We extend the proposed \gls{QCE} precoder to the case of QAM signaling. In particular, this is a novel extension of the constructive interference idea to non-PSK signals.
\end{enumerate}
\subsection{Remainder and Notation}
The remainder of this paper is organized as follows. In Section \ref{sec:sysmodel}, we present the system model. In Section \ref{sec:precoding_task}, the motivation behind formulating the precoding problem as a linear programming problem is explained. Sections \ref{sec:opt_problem_psk} and \ref{sec:opt_problem_qam} present the corresponding optimization problems for \gls{PSK} and \gls{QAM} signals, respectively. The complexity of each optimization problem is discussed in Section \ref{sec:complexity}. Simulation results are introduced in \ref{sec:simresults}. Finally, Section \ref{sec:conclusion} summarizes this work.

\textbf{Notation}: Bold lower case and upper case letters indicate vectors and matrices, non-bold letters express scalars. The operators $(.)^{*}$, $(.)^{\T}$ and $(.)^{\He}$ stand for complex conjugation, transposition and Hermitian transposition, respectively. The $n \times n$ identity (zero) matrix is denoted by $\matt{I}_{n}$ ($\matt 0_{n,n}$). The $n$ dimensional one (zero) vector is denoted by $\vect 1_n$ ($\vect 0_n$). The vector $\ve_m$ represents a zero-vector with $1$ at the $m$-th position. Additionally, $\diag\prths\va$ denotes a diagonal matrix containing the entries of the vector $\va$.
Every vector $\va$ of dimension $L$ is defined as $\va = \sum_{\ell=1}^L a_{\ell} \ve_l$. The operator $\otimes$ denotes the Kronecker product. The operator $\leq$ in the context of vector inequalities applies element-wise to the vector entries.
\section{System Model}
\label{sec:sysmodel}
\begin{figure*}[h]
\centering
\scalebox{1}{\tikzstyle{int}=[draw, minimum width=1cm, minimum height=1cm,  very thick]
\tikzstyle{init} = [pin edge={<-,thick,black}]
\tikzstyle{sum} = [draw, circle,inner sep=1pt, minimum size=2mm, very thick] 

\begin{tikzpicture}[node distance=1cm,auto,>=latex']
    \node [int] (a) {$\mathcal{P}\left(\bullet\right)$};
    \node (b) [left of=a,node distance=1.5cm, coordinate] {a};
    \node [int] (c) [right=1cm of a] {$\Q_{\text{CE}}\left(\bullet\right)$};
    \node [int] (e) [right=1cm of c] {$\matt{H}$};
    \node [sum,  pin={[init]below:$\boldsymbol{\eta}$}] (g) [right=1cm of e] {$\matt{+}$};
   
    \node [int] (i) [right=1cm of g] {$\matt{G}$};
     \node [int] (d) [right=1cm of i] {$\mathcal{D}\left(\bullet\right)$};
    \node [coordinate] (end) [right=1cm of d, node distance=1cm]{};

    \path[->,thick] (b) edge node[above] {$\vect{s}$} 
                       node[below] {$\mathbb{S}^{M}$} (a);
    \path[->,thick] (a) edge node[above] {$\vect{x}$}
                       node[below] {$\mathbb{X}^{N}$} (c);
    \path[->,thick] (c) edge node[above] {$\vect{t}$} 
                       node[below] {$\mathbb{T}^{N}$} (e);
    \path[->,thick] (e) edge node[above] {$\vect{y}$} 
                       node[below] {$\mathbb{C}^{M}$} (g);
    \path[->,thick] (g) edge node[above] {$\vect{r}$} 
                       node[below] {$\mathbb{C}^{M}$} (i); 
    \path[->,thick] (i) edge node[above] {$\vect{u}$} node[below] {$\mathbb{C}^{M}$}(d);                       
    \path[->,thick] (d) edge node[above] {$\hat{\vect{s}}$} 
                      node[below] {$\mathbb{S}^{M}$} (end) ;
\end{tikzpicture}}
\caption{Downlink MU-MIMO system model.}
\label{fig:model}
\end{figure*}
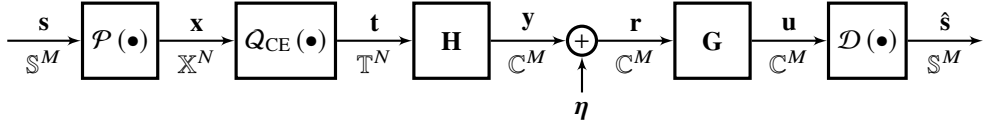
The system model shown in Fig.\ref{fig:model} consists of a single-cell massive \gls{MU}-\gls{MIMO} downlink scenario with coarsely QCE signals at the transmitter. The \gls{BS} is equipped with $N$ antennas and serves $M$ single-antenna users simultaneously, where $N\gg M$. The input signal vector $\vs$ contains the signals to be transmitted to each of the $M$ users. Each user's signal is drawn from the set $\mathbb{S}$ that represents either an $S$-\gls{PSK} or $S$-\gls{QAM} constellation, where $S$ denotes the number of constellation points. We assume that $\mathrm{E}[\vs]=\zeros_{M}$ and $\mathrm{E}[\vs\vs^{\He}]=\sigma^2_\text{s}\matt{I}_{M}$. The signal vector $\vs$ is precoded into the vector $\vx \in \mathbb{X}^N$ prior to the \glspl{DAC}. The non-linear function $\mathcal{P}\left(\bullet\right)$ is a symbol-wise precoder to reduce the distortions caused by the coarse quantization and the \gls{MUI}. The operator $\mathcal{Q}_{\text{CE}}(\bullet)$ models the non-linear behavior of the \glspl{DAC} combined with the power allocation at the \glspl{PA} as
\begin{align}
	\vt &= \mathcal{Q}_{\text{CE}}(\vx) = \sqrt{\frac{P_\text{tx}}{N}} \e^{\jim \mathcal{Q}_{{\phi}}(\arg(\vx))},
	\label{eq:t_from_x}
\end{align}
where the total transmit power $P_\text{tx}$ is allocated equally among the transmit antennas. The phase quantizer $\mathcal{Q}_{{\phi}}(\bullet)$ is a symmetric uniform real-valued quantizer. It is characterized by its resolution $q$ that defines the number of discrete output phases
\begin{align}
	Q = 2^q.
\end{align}
In other words,
the $2\pi$-phase range is divided into $Q$ $\frac{2\pi}{Q}$-rotationally symmetric sectors. The input signal that belongs to the $k$-th sector is quantized (mapped) to $\e^{\jim(2k-1)\frac{\pi}{Q}}$. This can be mathematically expressed as
\begin{align}
	\mathcal{Q}_{{\phi}}(\arg(x)) &= \left(\left\lfloor\frac{\arg(x)}{2\pi/Q}\right\rfloor+\frac{1}{2}\right) \frac{2\pi}{Q}.
\end{align}
Thus, the information after the \gls{CE} quantizer lies only in the phase. Hence, the set $\mathbb {T}$ is defined as
\begin{align}
\mathbb{T} = \left\lbrace \sqrt{\frac{P_\text{tx}}{N}} \exp\prths{\jim \prths{2i-1}\frac{\pi }{Q}}: i=1,\cdots,Q \right\rbrace.
\end{align}

The signal $\vt$ is transmitted through a flat-fading channel that is modeled by the channel matrix $\vH$. We assume that the $\left(m,n\right)$-th element $h_{mn}$ is a zero-mean unit-variance channel tap between the $n$-th transmit antenna and the $m$-th user. At the $M$ receive antennas, additive white Gaussian noise (AWGN), which is denoted by the vector $\boldsymbol{\eta} \sim \mathcal{C} \mathcal{N}_{\mathbb{C}}\left( \vect{0}_{M}, \mathbf{C}_{\boldsymbol{\eta}}=\matt{I}_{M}\right)$, perturbs the received signals 
\begin{align}
\vr = \vH\vt+\boldsymbol{\eta}.
\end{align}
Coherent data transmission with multiple \gls{BS} antennas
leads to an antenna gain, which depends on the channel realization. The entries of the received signal vector $\vr$ do not belong to the nominal decision regions of $\mathbb{S}$ but to a scaled version of them. Therefore, rescaling the received signal at each receive antenna is required to make the signal belong to the nominal decision region.  The rescaling operation is modeled by the diagonal real-valued matrix $\vG$, as follows
\begin{align}
\vu = \vG \prths{\vH \vt+\boldsymbol{\eta} },
\end{align}
where 
\begin{align}
\vG = \sum_{m=1}^M g_m \ve_m \ve_m^{\T},
\end{align}
with $g_m>0,~m=0,\cdots,M$. Note that no receive processing $\vG$ is required if $\mathbb{S}$ represents the \gls{PSK} constellation.
Finally, based on the decision regions to which the entries of the signal $\vu$ belong, the decision operation $\mathcal{D}(\bullet)$ produces the detected symbols $\hat{\vs}$  at the users
\begin{align}
\hat{\vs}=\mathcal{D}\left(\vG \left(\vH\vt+\boldsymbol{\eta} \right)\right).
\end{align}

\section{Precoding Task}
\label{sec:precoding_task}
In this work, we make use of the idea of constructive interference optimization \cite{Masouros.2013, Masouros.2015}.
When the downlink channel and all users' data are known at the transmitter, instantaneous constructive \gls{MUI} can be exploited to move the received signals further from the decision thresholds \cite{Masouros.2015}. In contrast to this, conventional precoding methods (MMSE, Zero-forcing) aim at minimizing the total \gls{MUI} such that the received signals lie as close as possible to the nominal constellation points. Constructive interference optimization exploits the larger symbol decision regions and thus leads to a more relaxed optimization. 

For every given input signal $\vs$ and for each channel realization $\vH$, the precoding task is to find
\begin{align}
\vx = \mathcal{P}\prths{\vs, \vH}.
\end{align} 
The task consists in designing the transmit vector $\vx$ such that $ \hat {\vs} = \vs$ holds true with high probability to reduce the detection error probability.
The symbol-wise precoder aims to mitigate all sources of distortion
\begin{itemize}
\item the quantization distortions
\item the channel distortions, and
\item the additive white Gaussian noise.
\end{itemize}
Our goal is to develop a problem formulation that jointly minimizes all three distortion sources. 

First, it is obvious that the quantization distortions can be omitted if we design the quantizer input such that it belongs to $\mathbb{T}^N$, i.e. $\mathbb{X}=\mathbb{T}$. Consequently, we would get an undistorted signal 
\begin{align}
\vt = \vx \text{, if } \vx \in \mathbb{T}^N.
\label{eq:QCE_constraint}
\end{align}
In what follows we enforce the QCE constraint in (\ref{eq:QCE_constraint}) to ensure the non-distorting behavior of the quantizer $\mathcal{Q}_{\text{CE}}(\bullet)$. 
\begin{figure}
\centering
\resizebox{0.3\textwidth}{!}{%
\begin{tikzpicture}

\path [fill, red!40] (2.1,0.2) -- (0.48,0.2) -- (2.1,1.82) ;
\path [fill, red!40] (0.2,2.1) -- (0.2,0.48) -- (1.82,2.1) ;

\path [fill, red!40] (-2.1,0.2) -- (-0.48,0.2) -- (-2.1,1.82) ;
\path [fill, red!40] (-0.2,2.1) -- (-0.2,0.48) -- (-1.82,2.1) ;

\path [fill, red!40] (2.1,-0.2) -- (0.48,-0.2) -- (2.1,-1.82) ;
\path [fill, red!40] (0.2,-2.1) -- (0.2,-0.48) -- (1.82,-2.1) ;

\path [fill, red!40] (-2.1,-0.2) -- (-0.48,-0.2) -- (-2.1,-1.82) ;
\path [fill, red!40] (-0.2,-2.1) -- (-0.2,-0.48) -- (-1.82,-2.1) ;

\draw [->, black] (-2.2,0) -- (2.2,0) node [below] {$\Re$};
\draw [->, black] (0,-2.2) -- (0,2.2) node [right] {$\jim \Im$};
\draw [dashed] (-2.1,-2.1) -- (2.1,2.1);
\draw [dashed] (2.1,-2.1) -- (-2.1,2.1);

\draw [<->, black] (0,1) -- (0.2,1) node [above,xshift=-0.08cm] {$\delta$};
%

\draw [fill,red] (0.9238795325,0.3826834324) circle (2pt);
\draw [fill,red] (-0.9238795325,0.3826834324) circle (2pt);
\draw [fill,red] (0.9238795325,-0.3826834324) circle (2pt);
\draw [fill,red] (-0.9238795325,-0.3826834324) circle (2pt);

\draw [fill,red] (0.3826834324,0.9238795325) circle (2pt);
\draw [fill,red] (0.3826834324,-0.9238795325) circle (2pt);
\draw [fill,red] (-0.3826834324,0.9238795325) circle (2pt);
\draw [fill,red] (-0.3826834324,-0.9238795325) circle (2pt);

\end{tikzpicture}
}
\caption{Decision and symbol regions (in red) for 8\gls{PSK} symbols.}
\label{fig:symbol_region_psk}
\centering
\resizebox{0.3\textwidth}{!}{%
\begin{tikzpicture}

\path [fill, red!40] (0.3162 - 0.1162,0.3162 - 0.1162) -- (0.3162 - 0.1162,0.3162 + 0.1162) -- (0.3162 + 0.1162,0.3162 + 0.1162) -- (0.3162 + 0.1162,0.3162 - 0.1162) ;

\path [fill, red!40] (0.9487 - 0.1162,0.9487 - 0.1162) -- (0.9487 - 0.1162,2.1) -- (2.1,2.1) -- (2.1,0.9487 - 0.1162) ;

\path [fill, red!40] (0.3162 + 0.1162,0.9487 - 0.1162) -- (0.3162 + 0.1162,2.1) -- (0.3162 - 0.1162,2.1) -- (0.3162 - 0.1162,0.9487 - 0.1162) ;

\path [fill, red!40] (0.9487 - 0.1162,0.3162 - 0.1162) -- (0.9487 - 0.1162,0.3162 + 0.1162) -- (2.1,0.3162 + 0.1162) -- (2.1,0.3162 - 0.1162) ;

\path [fill, red!40] (-0.3162 - 0.1162,0.3162 - 0.1162) -- (-0.3162 - 0.1162,0.3162 + 0.1162) -- (-0.3162 + 0.1162,0.3162 + 0.1162) -- (-0.3162 + 0.1162,0.3162 - 0.1162) ;

\path [fill, red!40] (-0.9487 + 0.1162,0.9487 - 0.1162) -- (-0.9487 + 0.1162,2.1) -- (-2.1,2.1) -- (-2.1,0.9487 - 0.1162) ;

\path [fill, red!40] (-0.3162 + 0.1162,0.9487 - 0.1162) -- (-0.3162 + 0.1162,2.1) -- (-0.3162 - 0.1162,2.1) -- (-0.3162 - 0.1162,0.9487 - 0.1162) ;

\path [fill, red!40] (-0.9487 + 0.1162,0.3162 - 0.1162) -- (-0.9487 + 0.1162,0.3162 + 0.1162) -- (-2.1,0.3162 + 0.1162) -- (-2.1,0.3162 - 0.1162) ;

\path [fill, red!40] (0.3162 - 0.1162,-0.3162 - 0.1162) -- (0.3162 - 0.1162,-0.3162 + 0.1162) -- (0.3162 + 0.1162,-0.3162 + 0.1162) -- (0.3162 + 0.1162,-0.3162 - 0.1162) ;

\path [fill, red!40] (0.9487 - 0.1162,-0.9487 + 0.1162) -- (0.9487 - 0.1162,-2.1) -- (2.1,-2.1) -- (2.1,-0.9487 + 0.1162) ;

\path [fill, red!40] (0.3162 + 0.1162,-0.9487 + 0.1162) -- (0.3162 + 0.1162,-2.1) -- (0.3162 - 0.1162,-2.1) -- (0.3162 - 0.1162,-0.9487 + 0.1162) ;

\path [fill, red!40] (0.9487 - 0.1162,-0.3162 - 0.1162) -- (0.9487 - 0.1162,-0.3162 + 0.1162) -- (2.1,-0.3162 + 0.1162) -- (2.1,-0.3162 - 0.1162) ;

\path [fill, red!40] (-0.3162 - 0.1162,-0.3162 - 0.1162) -- (-0.3162 - 0.1162,-0.3162 + 0.1162) -- (-0.3162 + 0.1162,-0.3162 + 0.1162) -- (-0.3162 + 0.1162,-0.3162 - 0.1162) ;

\path [fill, red!40] (-0.9487 + 0.1162,-0.9487 + 0.1162) -- (-0.9487 + 0.1162,-2.1) -- (-2.1,-2.1) -- (-2.1,-0.9487 + 0.1162) ;

\path [fill, red!40] (-0.3162 + 0.1162,-0.9487 + 0.1162) -- (-0.3162 + 0.1162,-2.1) -- (-0.3162 - 0.1162,-2.1) -- (-0.3162 - 0.1162,-0.9487 + 0.1162) ;

\path [fill, red!40] (-0.9487 + 0.1162,-0.3162 - 0.1162) -- (-0.9487 + 0.1162,-0.3162 + 0.1162) -- (-2.1,-0.3162 + 0.1162) -- (-2.1,-0.3162 - 0.1162) ;
\draw [->, black] (-2.2,0) -- (2.2,0) node [below] {$\Re$};
\draw [->, black] (0,-2.2) -- (0,2.2) node [right] {$\jim \Im$};

\draw [<->, black] (0,1.2) -- (0.3162 - 0.1162,1.2) node [above,xshift=-0.04cm] {$\delta$};

\draw [dashed] (-2.1,0.6325) -- (2.1,0.6325);
\draw [dashed] (-2.1,-0.6325) -- (2.1,-0.6325);
\draw [dashed] (0.6325,-2.1) -- (0.6325,2.1);
\draw [dashed] (-0.6325,-2.1) -- (-0.6325,2.1);

\draw [fill,red] (0.3162,0.33162) circle (2pt);
\draw [fill,red] (0.3162,-0.33162) circle (2pt);
\draw [fill,red] (-0.3162,0.33162) circle (2pt);
\draw [fill,red] (-0.3162,-0.33162) circle (2pt);

\draw [fill,red] (0.9487,0.9487) circle (2pt);
\draw [fill,red] (0.9487,-0.9487) circle (2pt);
\draw [fill,red] (-0.9487,0.9487) circle (2pt);
\draw [fill,red] (-0.9487,-0.9487) circle (2pt);

\draw [fill,red] (0.3162,0.9487) circle (2pt);
\draw [fill,red] (0.3162,-0.9487) circle (2pt);
\draw [fill,red] (-0.3162,0.9487) circle (2pt);
\draw [fill,red] (-0.3162,-0.9487) circle (2pt);

\draw [fill,red] (0.9487,0.3162) circle (2pt);
\draw [fill,red] (0.9487,-0.3162) circle (2pt);
\draw [fill,red] (-0.9487,0.3162) circle (2pt);
\draw [fill,red] (-0.9487,-0.3162) circle (2pt);
\end{tikzpicture}
}
\caption{Decision and symbol regions (in red) for 16\gls{QAM} symbols.}
\label{fig:symbol_region_qam}
\end{figure}

Second, to minimize the channel distortions and the noise, we look deeper at the constellation properties. As illustrated in Fig.~\ref{fig:symbol_region_psk} and Fig.~\ref{fig:symbol_region_qam}, each constellation is defined by the decision thresholds that separate the distinct decision regions of the constellation points. In total, we have as many contiguous decision regions as constellation points. Each constellation symbol lies within a \gls{SR} that is a downscaled version of the decision region. In contrast to the decision region, the \gls{SR} has a safety margin denoted by $\delta$ that separates it from the decision thresholds. When each entry of the noiseless received signal vector $\vect y$ belongs to the correct \gls{SR} and thus the correct decision region, the channel distortions are mitigated. Additionally, the safety margin $\delta$ has to be large enough such that the received signals when perturbed by the additive noise do not jump to the neighboring unintended decision regions. 

In summary, the problem formulation has to take into account the \gls{QCE} constraint, the \gls{SR} for each received signal and maximizing the safety margin. Thus, the optimization problem for the symbol-wise precoder can be written in general as follows
\begin{align}
&\max_{\vx}  \delta \label{eq:max_delta} \\
\text{ s.t.  }&  y_m \in \text{SR}_m, \forall m \label{eq:OR_constraint}\\
\text{and  }& \vx \in \mathbb{T}^N \label{eq:psk_constraint}.
\end{align}
This problem formulation depends on one hand on the input symbol vector $\vs$, which determines the intended \gls{SR} for each received signal, on the channel $\vH$, and on the other hand on the transmit power $P_{\text{tx}}$ that affects the noiseless received signal $\vy$. Since the optimization variables depend linearly on the square-root of the transmit power $\sqrt{P_{\text{tx}}}$, it is sufficient to solve the optimization problem for one transmit power value. Therefore, we consider the following specific case for the subsequent derivations:
\begin{align}
\vy' =  \vect y \mid_{P_\text{tx} =N, (\ref{eq:QCE_constraint})}= \vH \vx.
\label{eq:y'}
\end{align} 
This signal vector $\vy'$ is equal to the noiseless received signal $\vy$ for a transmit power $P_{\text{tx}} =N$ and when (\ref{eq:QCE_constraint}) is fulfilled. The optimization is based on this special case.
However, the \gls{QCE} constraint leads to a discrete optimization problem that cannot be solved efficiently. Therefore, the \gls{QCE} constraint will be relaxed to a convex constraint as shown in Section \ref{sec:relaxed_pol_constraint}. The constraint relaxation does not satisfy the equality in (\ref{eq:QCE_constraint}) and thus the quantization distortions are not fully omitted. However, they are reduced significantly as shown later.
\section{Problem Formulation for \gls{PSK} Signaling}
\label{sec:opt_problem_psk}
\subsection{Symbol Region for \gls{PSK} Signals}
In this section, we assume that the input signals $s_m$, $m=1, \cdots, M$, belong to the $S$-\gls{PSK} constellation. The set $\mathbb{S}$ in this case is defined as 
\begin{align}
\mathbb{S}:= \left\lbrace \exp\prths{\jim \prths{2i-1}\theta}: i=1,\cdots,Q \right\rbrace, ~\text{where}~\theta = \frac{\pi}{S}.
\end{align} 

\begin{figure}[h]
\centering  
\resizebox{0.3\textwidth}{!}{%
\scalefont{0.6}
\begin{tikzpicture}

\path [fill, red!40] (0.2828,0) -- (2.2,0.8) -- (2.2,-0.8);

\draw (0,-0.12) -- (2.2,0.8);
\draw (0,0.12) -- (2.2,-0.8);

\draw [->, black] (-0.5,0) -- (2.2,0) ;
\draw [->, black] (0,-0.5) -- (0,1);

\draw [fill,red] (1,0) circle (2pt) node [above,xshift=0.2cm] {$s_m$};
\draw [fill,blue] (1.7,0.5) circle (2pt) node [right,xshift=0.1cm] {$y'_m$};

\draw (1.7, 0.1) -- (1.7,-0.1) node [below] {$z_{m_R}$};
\draw (-0.1,0.5) -- (0.1,0.5) node [left, xshift=-0.1cm] { $z_{m_I}$};
\draw (0.2828,-0.1) -- (0.2828,0.1) node [below, yshift=-0.2cm] {$\tau$};
\draw [->, color=black] (0:0.65) arc (0:12:0.65) node [right,xshift=0.05cm] {$\theta$};
\end{tikzpicture}
}
\caption{Illustration of the \gls{PSK} symbol region in a modified coordinate system.}
\label{fig:opt_region_psk_mod_coordinates}
\end{figure}
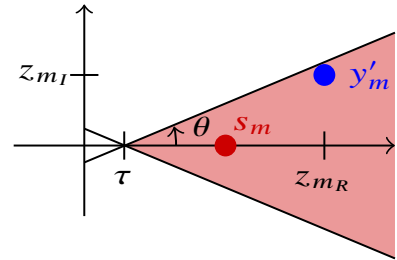

Each \gls{SR} in the \gls{PSK} constellation, as shown in Fig. \ref{fig:symbol_region_psk}, is a circular sector of infinite radius and angle $2\theta$. To find a mathematical expression for the \gls{SR}, the original coordinate system is rotated by the phase of the symbol of interest $s_m$ to get a modified coordinate system as illustrated in Fig.~\ref{fig:opt_region_psk_mod_coordinates}.
The coordinates of the noiseless received signal $y'_m$ in the modified coordinate system are given by
\begin{align}
z_{m_R} &= \Re\lbrace y_m' s_m^*\rbrace \frac{1}{\abs{ s_m}} \\
z_{m_I} &= \Im\lbrace y_m' s_m^*\rbrace \frac{1}{\abs{ s_m}} .
\end{align}
Since \gls{PSK} signals have unit magnitude, plugging (\ref{eq:y'}) into the above equations gives
\begin{align}
z_{m_R} &= \Re\lbrace \ve_m^{\T} \vH \vx s_m^*\rbrace = \Re\lbrace \ve_m^{\T} \matt {\tilde H} \vx \rbrace \label{eq:z_R} \\
z_{m_I} &= \Im\lbrace \vect e_m^{\T} \vH \vx s_m^*\rbrace = \Im\lbrace \vect e_m^{\T} \matt {\tilde H} \vx \rbrace,\label{eq:z_I}
\end{align}
where 
\begin{align}
\matt {\tilde H} = \diag(\vs^*) \vH.
\end{align}
The $m$-th \gls{SR} can be hence described by
\begin{align}
&z_{m_R} \geq \tau \label{eq:z_R_constraint} \\
&\vert z_{m_I} \vert \leq \left( z_{m_R} - \tau \right) \tan{\theta}, \forall m, \label{eq:z_I_constraint}
\end{align}
where $\tau = \frac{\delta}{\sin{\theta}}$. Note that the inequality in (\ref{eq:z_R_constraint}) is already fulfilled if the inequality in (\ref{eq:z_I_constraint}) is satisfied.
Plugging (\ref{eq:z_R}) and (\ref{eq:z_I}) into (\ref{eq:z_I_constraint}), the \glspl{SR} for all $M$ users can be defined by
\begin{align}
\vert \Im\lbrace \matt {\tilde H} \vx \rbrace\vert \leq \left( \Re\lbrace \matt {\tilde H}\vx \rbrace - \tau \vect 1_M\right) \tan{\theta}.
\end{align}
When using the following real-valued representation
\begin{align}
\Re\lbrace \matt {\tilde H}\vx \rbrace &= \underbrace{\begin{bmatrix}
\Re\lbrace \matt {\tilde H} \rbrace & -\Im\lbrace \matt {\tilde H} \rbrace \end{bmatrix}}_{=\matt A} \underbrace{\begin{bmatrix} \Re\lbrace \vx \rbrace\\ \Im\lbrace \vx \rbrace
\end{bmatrix}}_{= \vx'}= \matt A \vx' \\
\Im\lbrace \matt {\tilde H}\vx \rbrace &= \underbrace{\begin{bmatrix}
\Im\lbrace \matt {\tilde H} \rbrace & \Re\lbrace \matt {\tilde H} \rbrace \end{bmatrix}}_{=\matt B} \begin{bmatrix} \Re\lbrace \vx \rbrace\\ \Im\lbrace \vx \rbrace
\end{bmatrix} = \matt B \vx',
\end{align}
the constraint in (\ref{eq:OR_constraint}) can be rewritten as
\begin{align}
\begin{bmatrix} \matt B - \tan{\theta}\matt A &
    \frac{1}{\cos{\theta}} \vect 1_M \\ -\matt B - \tan{\theta}\matt A &
    \frac{1}{\cos{\theta}} \vect 1_M\end{bmatrix} \begin{bmatrix}\vx' \\ \delta \end{bmatrix}\leq \vect 0_{2M}.
    \label{eq:OR_linear_constraint}
\end{align}
\subsection{Relaxed Polygon Constraint}
\label{sec:relaxed_pol_constraint}
\begin{figure}
\centering
\resizebox{0.3\textwidth}{!}{%
\scalefont{0.6}
\begin{tikzpicture}

\draw [->, black] (-1.5,0) -- (1.5,0) node [below] {$\Re$};
\draw [->, black] (0,-1.5) -- (0,1.5) node [right] {$\jim \Im$};

\draw [fill,red] (0.9238795325,0.3826834324) circle (2pt);
\draw [fill,red] (-0.9238795325,0.3826834324) circle (2pt);
\draw [fill,red] (0.9238795325,-0.3826834324) circle (2pt);
\draw [fill,red] (-0.9238795325,-0.3826834324) circle (2pt);

\draw [fill,red] (0.3826834324,0.9238795325) circle (2pt);
\draw [fill,red] (0.3826834324,-0.9238795325) circle (2pt);
\draw [fill,red] (-0.3826834324,0.9238795325) circle (2pt);
\draw [fill,red] (-0.3826834324,-0.9238795325) circle (2pt);

\path [fill, green!40] (0.9238795325,0.3826834324) -- (0.9238795325,-0.3826834324) -- (0.3826834324,-0.9238795325)-- (-0.3826834324,-0.9238795325) -- (-0.9238795325,-0.3826834324) -- (-0.9238795325,0.3826834324) -- (-0.3826834324,0.9238795325) -- (0.3826834324,0.9238795325);

\end{tikzpicture}
}
\caption{Illustration of the relaxed polygon constraint for $Q$=8.}
\label{fig:pol_constraint}
\end{figure}
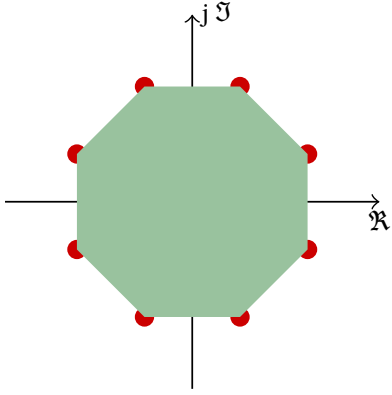

The non-convex \gls{QCE} constraint is relaxed to a convex constraint, which we call the polygon constraint, such that the entries of the vector $\vx$ belong to the polygon built by the $Q$-\gls{PSK} symbols, as shown in Fig.~\ref{fig:pol_constraint}. 
For the case of $Q=4$, we can describe the constraint as follows
\begin{align}
\vx' \leq \frac{1}{\sqrt{2}}\vect 1_{2N} \text{ and }
-\vx' \leq \frac{1}{\sqrt{2}} \vect 1_{2N}.
\end{align} 
For $q$-bit DACs, i.e., where the transmitted data are constrained to be $Q$-PSK symbols, the polygon can be constructed by the intersection of $Q/4$ squares that have an angular shift of $2\pi/Q$.
To this end, we define
\begin{align}
\matt T_i = \begin{bmatrix}
\cos{\beta_i} & \sin{\beta_i} \\ -\sin{\beta_i} & \cos{\beta_i}
\end{bmatrix} \otimes \matt I_N,~ i=1,...,Q/4,
\end{align}
where $ \beta_i = \frac{2\pi}{Q} (i-1)$.
The system of inequalities that considers the feasible set (polygon constraint) and hence relaxes the constraint in (\ref{eq:psk_constraint}) is given by
\begin{align}
\begin{bmatrix}
\matt T_1\\
-\matt T_1\\
\vdots \\
\matt T_{\frac{Q}{4}}\\
-\matt T_{\frac{Q}{4}}
\end{bmatrix} \vx' \leq \cos\left(\frac{\pi}{Q}\right) \vect 1_{NQ}.
\label{eq:relaxed_psk_constraint0}
\end{align}
Since $\vT_1=\vI_{2N}$, the first $4N$ inequalities in (\ref{eq:relaxed_psk_constraint0}) define the bounds of $\vx'$. Hence, (\ref{eq:relaxed_psk_constraint0}) can be rewritten as
\begin{align}
-\cos\left(\frac{\pi}{Q}\right) \vect 1_{2N} \leq ~&\vx' \leq \cos\left(\frac{\pi}{Q}\right) \vect 1_{2N},  \nonumber\\
\text{and}~~\underbrace{\begin{bmatrix}
\matt T_2\\
-\matt T_2\\
\vdots \\
\matt T_{\frac{Q}{4}}\\
-\matt T_{\frac{Q}{4}}
\end{bmatrix}}_{=\vE}~ &\vx' \leq \cos\left(\frac{\pi}{Q}\right) \vect 1_{NQ}.
\label{eq:relaxed_psk_constraint}
\end{align}
This reformulation leads to significant computational savings since the final optimization problem will be written as a linear program with bounded variables. It is beneficial in terms of computational complexity to have less number of inequalities as discussed in Section \ref{sec:complexity}. 
\subsection{Optimization Problem with the Relaxed Polygon Constraint}
Finally, the optimization problem for the symbol-wise precoder with \gls{PSK} signaling is obtained by combining (\ref{eq:max_delta}), (\ref{eq:OR_linear_constraint}) and (\ref{eq:relaxed_psk_constraint}) and is expressed as
\begin{align}
&\max_{\vv} \begin{bmatrix}
\vect 0_{2N}^{\T} & 1
\end{bmatrix} \vv \nonumber \\
&\text{ s.t. }  
\begin{bmatrix} \matt B - \tan{\theta}\matt A &
    \frac{1}{\cos{\theta}} \vect 1_M \\ -\matt B - \tan{\theta}\matt A &
    \frac{1}{\cos{\theta}} \vect 1_M \\ 
    \vE & \vect 0_{N\braces{Q-4}} 
     \end{bmatrix} \vv \leq \begin{bmatrix} \vect 0_{2M} \\ \cos\left(\frac{\pi}{Q}\right) \vect 1_{N\braces{Q-4}} \end{bmatrix},\nonumber \\
&\text{ and }    \mat{-\cos\left(\frac{\pi}{Q}\right) \vect 1_{2N}\\0}  \leq \vv \leq \mat{\cos\left(\frac{\pi}{Q}\right) \vect 1_{2N}\\\infty},
    \label{eq:final_optimization_problem_psk}
\end{align}
where $\vv^{\T} = \begin{bmatrix}
\vx'^{\T} & \delta
\end{bmatrix}$. 
The resulting optimization problem is a linear programming problem for which there exist very efficient solving methods \cite{Boyd.2004}.

When the optimization terminates, the optimal signal $\vx \in \mathbb{X}^{N}$ is found. The signal $\vt$ that goes through the channel is obtained as described in (\ref{eq:t_from_x}). In other words, each entry in $\vx$ gets mapped to the corresponding \gls{CE} point depending on the circular sector that it lies in.
\section{Problem Formulation for \gls{QAM} Signaling}
\label{sec:opt_problem_qam}
\subsection{The Need for an Additional Degree of Freedom $\alpha$}
In this section, we assume that the input signals $s_m$, $m=1, \cdots, M$, belong to the $S$-\gls{QAM} constellation, where $S$ is assumed to be a power of 4. The \gls{QAM} symbols are drawn from the set $\mathbb{S}$ defined as 
\begin{align}
\mathbb{S}:= \left\lbrace\pm \left(2i-1\right) \pm \jim \left(2i-1\right): i =1, \cdots, \log_4(S)  \right\rbrace.
\end{align} 
As explained in Section \ref{sec:precoding_task}, the safety margin $\delta$ has to be maximized such that the entries of the noiseless received signal $\vy$ belong to the intended \glspl{SR}. The \glspl{SR} in turn are determined by the constellation set $\mathbb{S}$ and the safety margin $\delta$. Hence, the safety margin $\delta$ cannot exceed $1$,
\begin{align}
\delta \leq 1.
\label{eq:delta_gamma}
\end{align}
Independently of the available transmit power, the entries of $\vy$ cannot have a distance to the decision thresholds larger than $1$. Hence, the available transmit power cannot be exploited to the fullest. 
This results already in a limitation of the problem formulation.

Thanks to the receive processing $\vG$, we can introduce an additional degree of freedom $\alpha$ such that the entries of the received signal $\vect y$ do not have to belong to the \glspl{SR} of the set $\mathbb{S}$ but rather to a scaled version of them; that is, the \gls{QAM} constellation at each receiver gets scaled by $ \alpha$. Thus, the constraint in (\ref{eq:delta_gamma}) is replaced by
\begin{align}
\delta \leq \alpha,
\label{eq:delta_alpha}
\end{align}
where $\alpha$ has to be jointly optimized with $\delta$.
Note that maximizing $\delta$ results in turn to maximizing $\alpha$, which leads to a maximal exploitation of the available transmit power. Thus, the entries of the signal vector $\vx$ will get closer to the polygon corners, which decreases the variations between $\vt$ and $\vx$. 

The ratio $\alpha$ denotes the expansion or shrinkage factor of the constellation at the receiver side depending on the available transmit power $P_{\text{tx}}$.
As explained in Section \ref{sec:precoding_task}, the optimization problem is formulated for the specific case, i.e. $P_{\text{tx}}=N$. 

\subsection{Scaled Symbol Region for \gls{QAM} Signals}
\label{sec:symbol_region_qam}
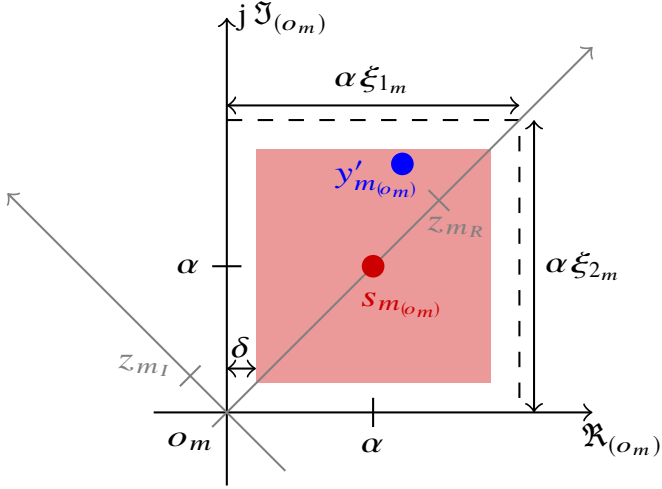
\begin{figure}
\centering
\resizebox{0.5\textwidth}{!}{%
\scalefont{0.6}
\begin{tikzpicture}

\draw [->, black] (-0.5,0) -- (2.5,0) node [below, xshift=0.2cm] {$\Re_{(o_m)}$};
\draw [->, black] (0,-0.5) -- (0,2.7) node [right] {$\jim \Im_{(o_m)}$};

\draw [<->, black] (0,0.3)--(0.2,0.3) node [above, xshift=-0.1cm] {$\delta$};
\draw (0,0) node [left, yshift=-0.2cm] { $o_m$};

\path [fill, red!40] (0.2,0.2) -- (1.8,0.2) -- (1.8,1.8) -- (0.2,1.8);
\draw [dashed] (0,2) -- (2,2) -- (2,0);

\draw [->, gray] (-0.1,-0.1) -- (2.5,2.5);
\draw [->, gray] (0.4,-0.4) -- (-1.5,1.5);
\draw [-,gray]({2.9/2+0.1/sqrt(2)},{2.9/2-0.1/sqrt(2)}) -- ({2.9/2-0.1/sqrt(2)},{2.9/2+0.1/sqrt(2)}) node [below,gray, xshift=0.2cm,yshift=-0.1cm] {$z_{m_R}$};
\draw [-,gray]({-0.1/sqrt(2)-0.5/2},{-0.1/sqrt(2)+0.5/2}) -- ({0.1/sqrt(2)-0.5/2},{0.1/sqrt(2)+0.5/2}) node [left,gray, xshift=-0.1cm] {$z_{m_I}$};

\draw [fill,red] (1,1) circle (2pt) node [below,xshift=0.2cm,yshift=-0.1cm] { $\tiny s_{m_{(\!o_m\!)}}$};
\draw [fill,blue] (1.2,1.7) circle (2pt) node [left,xshift=0.2cm,yshift=-0.1cm] { $\tiny y'_{m_{(\!o_m\!)}}$};

\draw (1,-0.1) -- (1,0.1) node [below, yshift=-0.2cm] { $\alpha$};
\draw (-0.1,1) -- (0.1,1) node [left, xshift=-0.2cm] { $\alpha$};

\draw [<->,black] (0,2.1) -- (2,2.1) node [above, xshift=-1cm] { $\alpha \xi_{1_m}$};

\draw [<->,black] (2.1,0) -- (2.1,2) node [right, yshift=-1cm] { $\alpha \xi_{2_m}$};

\end{tikzpicture}
}
\caption{Illustration of the \gls{QAM} receiver symbol region for $\Re\left\lbrace s_m \right\rbrace > 0$ and $\Im\left\lbrace s_m \right\rbrace > 0$ in the shifted coordinate system (in black) and in the shifted and rotated coordinate system (in gray): $\xi_{{1/2}_m} \in \lbrace2,\infty\rbrace$.}
\label{fig:opt_region_qam}
\end{figure}

To describe the \glspl{SR} for \gls{QAM} signaling after considering $\alpha$, we define a new coordinate system, that is a shifted and rotated version of the original coordinate system. 
First, the original receiver constellation system is shifted by $o_m$
\begin{align}
o_m &= \alpha \Bigg(\left(\Re\left\lbrace s_m \right\rbrace -\sign\left( \Re\left\lbrace  s_m \right\rbrace \right)\right)\nonumber \\
&\qquad +\jim \left(\Im\left\lbrace s_m  \right\rbrace -\sign\left( \Im\left\lbrace s_m \right\rbrace \right)\right)\Bigg).
\label{eq:o_m}
\end{align} 
We get the following expressions for the received and the desired signal in the new coordinate system depicted in Fig.~\ref{fig:opt_region_qam}
\begin{align}
{y_m}'_{(o_m)}  &= y_m' - o_m \nonumber \\
&= \vect e_m^{\T} \vH \vx - o_m \label{eq:y_m}\\
{s_m}_{(o_m)}  &= \alpha s_m - o_m \nonumber \\
&= \alpha \left(\sign\left( \Re\left\lbrace s_m \right\rbrace \right)+\jim \sign\left( \Im\left\lbrace s_m \right\rbrace \right) \right)\label{eq:s_m}.
\end{align}

Second, the intermediate coordinate system is rotated by the phase of the symbol of interest ${s_m}_{(o_m)}$. So the received signal $y_m'$ has the following coordinates in the shifted and rotated coordinate system
\begin{align}
z_{m_R}&=\frac{\Re\left\lbrace {y_m}'_{(o_m)}{s_m}_{(o_m)}^* \right\rbrace}{\vert {s_m}_{(o_m)}\vert},
\label{eq:real_projection_qam}
\end{align}
and
\begin{align}
z_{m_I}&=\frac{\Im\left\lbrace {y_m}'_{(o_m)}{s_m}_{(o_m)}^* \right\rbrace}{\vert {s_m}_{(o_m)}\vert}.
\label{eq:imag_projection_qam}
\end{align}
We get
\begin{align}
\frac{{y_m}'_{(o_m)}{s_m}_{(o_m)}^*}{\vert {s_m}_{(o_m)}\vert} &= \frac{1}{\sqrt{2}\alpha}\left(\vect e_m^{\T} \vH \vx - o_m \right){s_m}_{(o_m)}^* \nonumber \\
&=\frac{1}{\sqrt{2}\alpha} \left({s_m}_{(o_m)}^*\vect e_m^{\T} \vH \vx - o_m{s_m}_{(o_m)}^*\right) \nonumber \\
&= \frac{\left(\sign\left( \Re\left\lbrace s_m \right\rbrace \right)-\jim \sign\left( \Im\left\lbrace s_m \right\rbrace \right) \right)}{\sqrt{2}} \vect e_m^{\T} \vH \vx  \nonumber \\
& \quad \qquad\qquad \qquad\ \qquad - \frac{1}{\sqrt{2}\alpha} o_m{s_m}_{(o_m)}^* \nonumber \\
& = \vect e_m^{\T}  {\hat{\vH}} \vx- \alpha c_m,
\label{eq:projection_term_qam}
\end{align}
where 
\begin{align}
\hat{\vH} &=  \frac{1}{\sqrt{2}}\diag\left(\sign\left( \Re\left\lbrace \vs \right\rbrace \right)-\jim \sign\left( \Im\left\lbrace \vs \right\rbrace \right) \right) \vH,
\label{eq:hat_H_1}
\end{align}
and 
\begin{align}
c_m &=\frac{o_m{s_m}_{(o_m)}^*}{\sqrt{2}\alpha^2}.
\label{eq:c_m}
\end{align}
Note that $c_m$ does not actually depend on $\alpha$ as can be concluded from (\ref{eq:o_m}), (\ref{eq:s_m}) and (\ref{eq:c_m}).
Plugging (\ref{eq:projection_term_qam}) into (\ref{eq:real_projection_qam}) and (\ref{eq:imag_projection_qam}), we get 
\begin{align}
z_{m_R} &= \vect e_m^{\T} \matt V \vx' - \alpha\Re\left\lbrace c_m\right\rbrace \\
z_{m_I} &= \vect e_m^{\T} \matt W \vx' - \alpha\Im\left\lbrace c_m\right\rbrace, 
\end{align}
where 
\begin{align}
\matt V & = \begin{bmatrix}
\Re\lbrace \matt {\hat H} \rbrace & -\Im\lbrace \matt {\hat H} \rbrace \end{bmatrix} \label{eq:V}\\
\matt W & = \begin{bmatrix}
\Im\lbrace \matt {\hat H} \rbrace & \Re\lbrace \matt {\hat H} \rbrace \end{bmatrix} \label{eq:W}.
\end{align}
The $m$-th \gls{SR}, as shown in Fig.~\ref{fig:opt_region_qam}, can be hence described by
\begin{align}
&z_{m_R} \geq \sqrt{2} \delta  \label{eq:symbol_region_qam_constraint1}\\
&z_{m_R} \leq \sqrt{\left(\alpha \xi_{1_m}-\delta\right)^2 + \left(\alpha \xi_{2_m}-\delta\right)^2} \label{eq:symbol_region_qam_constraint2} \\
&\vert z_{m_I}\vert \leq \left( z_{m_R} - \sqrt{2} \delta \right)\label{eq:symbol_region_qam_constraint3}\\
& z_{m_I} \leq -z_{m_R} + \sqrt{2}\left(\alpha \xi_{2_m}-\delta \right)\label{eq:symbol_region_qam_constraint4} \\
& z_{m_I} \geq z_{m_R} - \sqrt{2}\left(\alpha\xi_{1_m}-\delta \right)\label{eq:symbol_region_qam_constraint5}.
\end{align}
Note that $\xi_{1_m}$ and $\xi_{2_m} \in \lbrace 2,\infty \rbrace$ depending on which constellation point the symbol of interest $s_m$ corresponds to. If $s_m$ is one of the outer constellation points, then at least $\xi_{1_m}$ or $\xi_{2_m}$ must be equal to $\infty$. Since (\ref{eq:symbol_region_qam_constraint1}) and (\ref{eq:symbol_region_qam_constraint2}) are inherently fulfilled by (\ref{eq:symbol_region_qam_constraint3}), (\ref{eq:symbol_region_qam_constraint4}) and (\ref{eq:symbol_region_qam_constraint5}), the constraint in (\ref{eq:OR_constraint}) can be rewritten as
\begin{align}
 \begin{bmatrix}
\matt W-\matt V \! &\! \vect 1_M \!&\!  \Re\left\lbrace \vc \right\rbrace -\Im\left\lbrace \vc \right\rbrace\\
-\matt W-\matt V \! &\!  \vect 1_M \!&\! \Re\left\lbrace \vc \right\rbrace +\Im\left\lbrace \vc \right\rbrace\\
\matt W+\matt V  \!&\!  \vect 1_M \!&\! - \Re\left\lbrace \vc \right\rbrace -\Im\left\lbrace \vc \right\rbrace-\sqrt{2} \boldsymbol \xi_2\\
-\matt W+\matt V \! &\!  \vect 1_M \!&\! - \Re\left\lbrace \vc \right\rbrace +\Im\left\lbrace \vc \right\rbrace-\sqrt{2} \boldsymbol \xi_1 \\
\end{bmatrix} \!\!\begin{bmatrix}
\vx'\\ \sqrt{2} \delta \\ \alpha
\end{bmatrix} \!\! \leq 
\matt 0_{4M}.
\label{eq:sr_constraint_qam}
\end{align}

\subsection{Optimization Problem with the Relaxed Polygon Constraint}
We are interested in maximizing the safety margin as presented in (\ref{eq:max_delta}). In contrast to the \gls{PSK} case, there is a constraint on $\delta$ in the \gls{QAM} case, stated in (\ref{eq:delta_alpha}), which is inherently fulfilled by (\ref{eq:sr_constraint_qam}).
Combining (\ref{eq:max_delta}) with the \gls{SR} constraint in (\ref{eq:sr_constraint_qam}) and the relaxed polygon constraint in (\ref{eq:relaxed_psk_constraint}), we get a linear programming problem for the design of the symbol-wise precoder for \gls{QAM} signaling. The optimization problem is given in (\ref{eq:final_optimization_problem_qam}),
\begin{figure*}[t] 
\begin{align}
\max_{\vv} \begin{bmatrix}
\vect 0_{2N}^{\T} & 1 & 0
\end{bmatrix} \vv &\text{ s.t. }  \begin{bmatrix}
\matt W-\matt V \! &\! \vect 1_{M} \!&\!  \Re\left\lbrace \vc \right\rbrace -\Im\left\lbrace \vc \right\rbrace \\
-\matt W-\matt V \! &\!  \vect 1_{M} \!&\! \Re\left\lbrace \vc\right\rbrace +\Im\left\lbrace \vc \right\rbrace\\
\matt W+\matt V  \!&\!  \vect 1_{M} \!&\! - \Re\left\lbrace \vc \right\rbrace -\Im\left\lbrace \vc \right\rbrace-\sqrt{2} \boldsymbol \xi_{2}\\
-\matt W+\matt V \! &\!  \vect 1_{M} \!&\! - \Re\left\lbrace \vc \right\rbrace +\Im\left\lbrace \vc \right\rbrace-\sqrt{2} \boldsymbol \xi_{1} \\
\vE \!&\! \matt 0_{N\braces{Q-4}}\!&\! \matt 0_{N\braces{Q-4}}
\end{bmatrix} \vv \leq  \begin{bmatrix}
\matt 0_{4M}\\
\cos\left( \frac{\pi}{Q}\right) \matt 1_{N\braces{Q-4}}
\end{bmatrix}\nonumber \\
&\text{and }   \mat{-\cos\left(\frac{\pi}{Q}\right) \vect 1_{2N}\\0\\ 0}  \leq \vv \leq \mat{\cos\left(\frac{\pi}{Q}\right) \vect 1_{2N}\\\infty \\ \infty}.
\label{eq:final_optimization_problem_qam}
\end{align}
\end{figure*}
where $\vv^{\T} = \begin{bmatrix}
\vx'^{\T} & \sqrt{2}\delta & \alpha
\end{bmatrix}$.

Again the optimized vector $\vx \in \mathbb{X}^N$ goes through the quantizer, as stated in (\ref{eq:t_from_x}), to obtain the transmit vector $\vt$.
\subsection{Receive Processing}
The variables of the optimization problem are the transmit vector $\vx$, the safety margin $\delta$ and the expansion factor $\alpha$. The latter determines the receive processing $\matt G$. Note that the optimal value of $\alpha$ is determined on a symbol-by-symbol basis, and its value cannot be communicated to the receiver. However, due to the massive \gls{MIMO} assumption and the induced hardening effect, the flactuations of $\alpha$ across the symbols are small (see also the discussion in the following subsections). Therefore an exact value of $\alpha$ is not required at the receiver. Only the positions of the decision thresholds are needed to rescale the receiver constellation points to the nominal constellation points, and these only depend on the mean value of $\alpha$. An estimate of the mean of $\alpha$ can easily be computed by averaging over a block of received signals.

After multiplication with the receiver coefficient $g_m$, the scaled received signal should equal
\begin{align}
u_m &= g_m r_m = g_m \e_m^{\T} \vH \vt + g_m \eta_m  = s_m + \eta'_m,
\end{align}
where $\eta'_m$ denotes the deviation of $u_m$ from the nominal point $s_m$ due to the \gls{AWGN} $\eta_m$, the \gls{SR} constraint and the quantization applied on the relaxed optimized vector $\vx$. Then, we can write
\begin{align}
|{\rm Re}\{r_m\}|\!+\!|{\rm Im}\{r_m\}|\!\!=& g_m^{-1} \braces{|{\rm Re}\{\!s_m\!+\! \eta'_m\}|\!\!+\!\! |{\rm Im}\{\!s_m\!+\! \eta'_m\}|} \\
 \stackrel{\textrm{w/ high prob. at SINR $\gg$ 1}}{=}&g_m^{-1}  \left(|{\rm Re}\{s_m\}| +|{\rm Im}\{s_m\}|\right) \nonumber \\
 +&\: g_m^{-1} \braces{ {\rm Re}\{\eta'_m\} +{\rm Im}\{\eta'_m\}},
\end{align}
meaning that with zero-mean noise plus interference $\eta'_m$ we have
\begin{align}
{\rm E}[|{\rm Re}\{r_m\}|+|{\rm Im}\{r_m\}| ] \approx g_m^{-1} {\rm E}[|{\rm Re}\{s_m\}|+|{\rm Im}\{s_m\}|].
\label{eq:g_explanation}
\end{align} 
Based on (\ref{eq:g_explanation}), we propose a blind estimation method to obtain the scaling factor $g_m$ for each user prior to the decision operation. The method does not require any feedback or training from the \gls{BS} nor any knowledge of the noise plus interference power at the user terminal:
\begin{equation}
g_m= T \cdot \frac{{\rm E}\left[|{\rm Re}\{s\}|+|{\rm Im}\{s\}|\right]}{ \sum_{t=1}^T  |{\rm Re}\{r_m[t]\}|+|{\rm Im}\{r_m[t]\}|},
\end{equation}
where $T$ is  the length the received sequence. 

\subsection{Symbol-wise Processing vs. Block-wise Processing}
\begin{figure}
\centering
\begin{minipage}{0.05\textwidth}
\vspace{-1cm}
\small $B=1$\\

\vspace{2.5cm}

\small $B=4$
\end{minipage}%
\begin{minipage}{0.5\textwidth}
\begin{flushleft}
\includegraphics[scale=0.6]{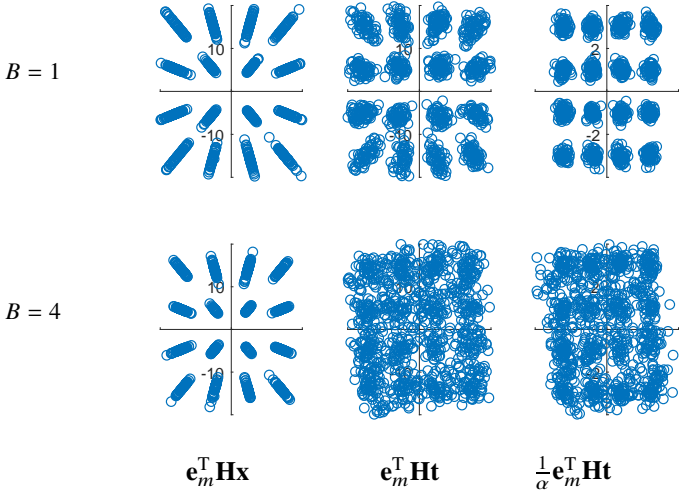}
\end{flushleft}
\vspace{-0.5cm}
\hspace{1.5cm}$\ve_m^{\T}\vH\vx$ \hspace{1.5cm} $\ve_m^{\T}\vH\vt$ \hspace{1cm} $\frac{1}{\alpha}\ve_m^{\T}\vH\vt$ 
\end{minipage}
\caption{The noiseless received symbols at one arbitrary user $m$ for an arbitrary \gls{iid} channel realization with $N=64$, $M=8$ and $Q=4$.}
\label{fig:effectofB}
\end{figure}

One might ask why we opt for symbol-wise processing and not block-wise processing. The factor $\alpha$ cannot be communicated to the receiver and hence has to be estimated. The estimation is based on averaging over a block of $T$ received signals. Thus, one expects that the design of $\alpha$ at the transmitter has to be computed for the same block length, i.e. $B=T$. However, fixing $\alpha$ for a certain block length not only increases the complexity of the problem but also reduces the degrees of freedom of the optimization problem at the transmitter and the vectors $\vx$ have to be designed with a greater restriction on $\alpha$. This leads to the entries of the vector $\vx$ moving farther from the polygon corners, thus increasing the quantization distortions. This effect is illustrated in Fig.~\ref{fig:effectofB}, where the entries of $\ve_m^{\T}\vH \vx$, $\ve_m^{\T}\vH \vt$ and $\frac{1}{\alpha}\ve_m^{\T}\vH \vt$ of an arbitrary user $m$ are obtained by transmitting 1024 16QAM signal vectors through an \gls{iid} channel of $N=64$, $M=8$ and $Q=4$. The optimization is computed for both symbol-wise processing, i.e. $B=1$, and block-wise processing with $B=4$. As can be deduced from the plots, the block-wise processing leads to a larger safety margin with the relaxed vector $\vx$. However, after applying the quantization this gain is lost and the symbol-wise processing seems to be more robust against the quantization operation. This can be further explained by the results in Table \ref{tab:mse_B}, which shows $\E\sqbrackets{\frac{\norm{\vt \neq \vx}_1}{N}}$, the percentange of entries of $\vx$ that are distorted due to the quantization and the \gls{MSE} between $\vt$ and $\vx$. We see that increasing $B$ significantly increases the quantization distortion.
\begin{table}[h!]
  \begin{center}
    \caption{Quantization distortion vs. $B$.}
    \label{tab:mse_B}
    \begin{tabular}{|c|c|c|} 
 	  \hline
        & $B=1$ & $B=4$ \\
      \hline 
      $\E\sqbrackets{\frac{\norm{\vt \neq \vx}_1}{N}}$  & 0.2176 & 0.4432 \\
      \hline
      $\E\sqbrackets{\norm{\vt-\vx}^2_2}$  & 2.5458 & 12.6429 \\
      \hline
    \end{tabular}
  \end{center}
\end{table}
Therefore, the symbol-wise processing is chosen in this contribution, i.e, an optimal value of $\alpha$ is designed for each vector $\vx$. 
\subsection{One Joint $\alpha$ vs. $M$ Distinct $\alpha$'s for $M$ Users}
The symbol-wise transmit processing followed by the block-wise receive processing is reliable only if the obtained values of $\alpha$ do not vary much from one vector $\vx$ to another. Otherwise, estimating the mean value of $\alpha$ at the receiver would not be sufficient for correct detection.
This explains why we choose one joint $\alpha$ for all users. If a different value $\alpha_m$ per user is chosen, this would result in more degrees of freedom and the values $\alpha_m,m=1,\cdots,M$, would fluctuate much more from one vector $\vx$ to another, which worsens the estimation result at the receiver. For a large number of users, the jointly designed $\alpha$ will not vary much due to the channel hardening effect. This behavior can be better illustrated by looking at the relative range of $\alpha$
\begin{align}
{\E}_{\vH}\sqbrackets{\frac{\max\{\alpha\} - \min\{\alpha\}}{\E\sqbrackets{\alpha}}}
\end{align}
and the maximal relative range of $\alpha_m,~m=1,\cdots,M$,
\begin{align}
\max_m {\E}_{\vH}\sqbrackets{\frac{\max\{\alpha_m\} - \min\{\alpha_m\}}{\E\sqbrackets{\alpha_m}}}
\end{align}
averaged over many channel realizations and for different values of $N$ and $M$ in Table \ref{tab:range_alpha} and Table \ref{tab:range_alpham} with 16QAM signaling and \gls{iid} channel. 
\begin{table}[h!]
  \begin{center}
    \caption{Relative Range of $\alpha$: ${\E}_{\vH}\sqbrackets{\frac{\max\{\alpha\} - \min\{\alpha\}}{\E\sqbrackets{\alpha}}}$.}
    \label{tab:range_alpha}
    \begin{tabular}{|l|c|c|} 
 	  \hline
       \backslashbox{$M$}{$N$} & $64$ & $200$ \\
      \hline 
       $2$& 1.52 & 1.44 \\
      \hline
      $8$& 0.78 & 0.71 \\
      \hline
      $14$& 0.61 &0.50 \\
      \hline
    \end{tabular}
  \end{center}
\end{table}
\begin{table}[h!]
  \begin{center}
    \caption{Maximal Relative Range of $\alpha_m$: $\max\limits_m {\E}_{\vH}\sqbrackets{\frac{\max\{\alpha_m\} - \min\{\alpha_m\}}{\E\sqbrackets{\alpha_m}}}$.}
    \label{tab:range_alpham}
    \begin{tabular}{|l|c|c|} 
 	  \hline
      \backslashbox{$M$}{$N$}  & $64$ & $200$ \\
      \hline 
       $2$& 1.51 & 1.44 \\
      \hline
      $8$& 1.37 & 0.75 \\
      \hline
      $14$& 2 &0.97 \\
      \hline
    \end{tabular}
  \end{center}
\end{table}
As a consequence, exploiting the channel hardening effect by using only one single $\alpha$ for all users is crucial for the robustness of the method and to allow for adequate receiver processing for calculating the scaling factors. 

\section{Computational Complexity of MSM}
\label{sec:complexity}
\subsection{On the Computational Complexity of General Linear Programming Problems}
In this section, we study the computational complexity of the simplex method for a general linear programming problem with bounded variables in inequality form:
\begin{align}
\max_{\vx} \vc^{\T} \vx ~&\text{s.t.}~\vA \vx \leq \vb \nonumber \\
& \text{and}~ \vl\leq\vx \leq\vu,
\end{align}
where $\vc,~\vx,~\vl$ and $\vu \in \mathbb{R}^{n}$, $\vA\in \mathbb{R}^{m \times n}$ and $\vb \in \mathbb{R}^{m}$. 

First, we have to make sure that the entries of $\vb$ are non-negative. To this end, we change the signs of the inequalities that correspond to negative entries in $\vb$. So we get
\begin{align}
\min_{\vx } \vc^{\T} \vx ~&\text{s.t.}~\tilde{\vA} \vx ~\substack{\geq \\ \leq}~ \tilde{\vb}  \nonumber \\
& \text{and}~ \vl\leq\vx  \leq\vu,
\label{eq:lp_inequality}
\end{align}
where $\tilde{\vb} \in \mathbb{R}^{m}_{+}$ and some inequalities hold with the sign $\leq$ and others with the sign $\geq$.

Second, the linear programming problem is transformed to the canonical form by introducing $m$ slack and surplus variables denoted by $\vx_{\text{s}}$. Additionally, $a$ artificial variables denoted by $\vx_{\text{a}}$, with $0\leq a \leq m$, are added to set up an initial feasible solution \cite{Dantzig.1997}. The equivalent enlarged problem reads as
\begin{align}
\min_{\bar{\vx} } \bar{\vc}^{\T} \bar{\vx} ~&\text{s.t.}~\bar{\vA} \bar{\vx} = \tilde{\vb}  \nonumber \\
& \text{and}~ \bar{\vl}\leq\bar{\vx}  \leq\bar{\vu},
\label{eq:lp_bounded}
\end{align}
where $\bar \vA = \mat{\tilde{\vA} & \vA_{\text{s}} & \vI_{a}}\in \mathbb{R}^{m \times (n+m+a)}$, $\bar{\vx}^{\T}=\mat{\vx^{\T} & \vx_{\text{s}}^{\T} & \vx_{\text{a}}^{\T}}\in \mathbb{R}^{n+m+a}$, $\bar{\vl}^{\T} = \mat{\vl^{\T} &  \zeros_{m+a}^{\T}}$ and $\bar{\vu}^{\T} = \mat{\vu^{\T}& \infty \ones_{m+a}^{\T}}$. The matrix $\vA_{\text{s}}$ is a diagonal matrix with entries equal to $1$ or $-1$ depending on whether the inequality sign in (\ref{eq:lp_inequality}) is $\leq$ or $\geq$, respectively. The number $a$ of artificial variables is defined by the number of negative entries in $\vA_{\text{s}}$, such that the concatenation of $m$ columns from $\mat{\vA_{\text{s}} & \vI_{a}}$ can construct the identity matrix $\vI_{m}$. For the special case $\vb = \tilde \vb$, i.e. the entries of $\vb$ are non-negative, $\vA_{\text{s}} = \vI_{m}$. Hence, no artificial variables are needed, i.e. $a=0$.

With the use of the simplex method to solve (\ref{eq:lp_bounded}), the number of operations (multiplication and addition pairs) on each iteration is given by, \cite[p.83]{Dantzig.1997},
\begin{align}
3m \qquad \text{or} \qquad (m+1)(n+a+1)+ 2m,
\end{align}
depending on whether pivoting is required or not. According to \cite[p.86]{Dantzig.1997}, in most iterations no pivoting is required and hence less computation is needed.

\subsection{Computational Complexity of MSM for PSK Signaling}
As can be seen from (\ref{eq:final_optimization_problem_psk}), there are $m=2M+N\braces{Q-4}$ inequalities and $n=2N+1$ variables. The number $a$ of artificial variables reduces to $0$, since the vector $\vb^{\T}=\mat{\zeros_{2M}^{\T} & \cos\left(\frac{\pi}{Q}\right) \vect 1_{N\braces{Q-4}}^{\T}}$ has only non-negative entries. Thus, the number of operations (multiplication and addition pairs) on each iteration calculates in this case to
\begin{align}
6M+3\braces{Q-4}N,
\end{align}
or
\begin{align}
2N+4MN+8M+2\braces{Q-4}\braces{2N^2+4N}.
\end{align}
For the special case of 1-bit quantization, i.e. $Q=4$, the complexity is linear in $N$ and $M$.

\subsection{Computational Complexity of MSM for QAM Signaling}
From (\ref{eq:final_optimization_problem_qam}), we have  $m=4M+N\braces{Q-4}$ inequalities and $n=2N+2$ variables. The number $a$ of artificial variables reduces to $0$, since the vector $\vb^{\T}=\mat{\zeros_{4M+1}^{\T} & \cos\left(\frac{\pi}{Q}\right) \vect 1_{N\braces{Q-4}}^{\T}}$ has only non-negative entries. Thus, the number of operations (multiplication and addition pairs) on each iteration calculates in this case to
\begin{align}
12M+3\braces{Q-4}N,
\end{align}
or
\begin{align}
2N+8MN+20M+3+\braces{Q-4}\braces{2N^2+5N}.
\end{align}
For the special case of 1-bit quantization, i.e. $Q=4$, the complexity is linear in $N$ and $M$. Note that the sparsity of $\vE$ can be exploited by deploying the revised simplex method to reduce the number of required operations in the case of $Q>4$\cite[p.89]{Dantzig.1997}.

\section{Simulation Results}
\label{sec:simresults}
For the simulations, we assume a
\gls{BS} with $N = 64$ antennas serving $M = 8$ single-antenna
users. The channel $\vH$ is composed of \gls{iid} Gaussian
random variables with zero-mean and unit variance. 
The numerical results are obtained with Monte Carlo simulations of 100 independent channel realizations.
The additive
noise is also \gls{iid} with variance one at each antenna. 
The performance metric is the uncoded \gls{BER} averaged over the single-antenna users. For the blind estimation of the coefficients $g_m$ we use a block length of $T=128$.

\begin{figure}
\centering
\resizebox{0.5\textwidth}{!} {
%
%
\definecolor{mycolor1}{rgb}{0.00000,0.45098,0.74118}%
\definecolor{mycolor2}{rgb}{0.47059,0.67059,0.18824}%
\definecolor{mycolor3}{rgb}{0.63922,0.07843,0.18039}%
\definecolor{mycolor4}{rgb}{0.92941,0.69020,0.12941}%
\definecolor{mycolor5}{rgb}{0.85098,0.32941,0.10196}%
\definecolor{mycolor6}{rgb}{0.49020,0.18039,0.56078}%
\begin{tikzpicture}[spy using outlines={rectangle,lens={scale=2}, size=0.8cm, connect spies}]

\begin{axis}[%
width=\columnwidth,
height=0.25\textheight,
at={(0.758in,0.481in)},
scale only axis,
xmin=-6,
xmax=12,
xlabel={$P_{\text{tx}}$~(dB)},
xmajorgrids,
ymode=log,
ymin=1e-4,
ymax=1,
yminorticks=true,
ylabel={Uncoded BER},
ymajorgrids,
yminorgrids,
axis background/.style={fill=white},
title style={font=\bfseries},
legend style={legend cell align=left,align=left,draw=white!15!black}
]

\addplot [color=mycolor1,solid,line width=2.0pt,mark=+,mark options={solid},mark size=3pt]
   table[row sep=crcr]{%
-10	0.253863525390625\\
-8	0.206170654296875\\
-6	0.1484375\\
-4	0.095928955078125\\
-2	0.05191650390625\\
0	0.021954345703125\\
2	0.0060791015625\\
4	0.001043701171875\\
6	6.7138671875e-05\\
8	0\\
10	0\\
12	0\\
14	0\\
16	0\\
18	0\\
20	0\\
22	0\\
24	0\\
26	0\\
28	0\\
};
\addlegendentry{MSM};

\addplot [color=mycolor2,solid,line width=2.0pt,mark=triangle,mark options={solid,rotate=270}]
  table[row sep=crcr]{%
-10	0.242449951171875\\
-8	0.19512939453125\\
-6	0.139056396484375\\
-4	0.089013671875\\
-2	0.047760009765625\\
0	0.019952392578125\\
2	0.005694580078125\\
4	0.000982666015625\\
6	6.103515625e-05\\
8	0\\
10	0\\
12	0\\
14	0\\
16	0\\
18	0\\
20	0\\
22	0\\
24	0\\
26	0\\
28	0\\
};
\addlegendentry{SQUID};

\addplot [color=mycolor3,solid,line width=2.0pt,mark=square,mark options={solid}]
table[row sep=crcr]{%
-10	0.242889404296875\\
-8	0.19505615234375\\
-6	0.143109130859375\\
-4	0.0973876953125\\
-2	0.05806884765625\\
0	0.030560302734375\\
2	0.014593505859375\\
4	0.006689453125\\
6	0.0028076171875\\
8	0.001513671875\\
10	0.000958251953125\\
12	0.000567626953125\\
14	0.000390625\\
16	0.000286865234375\\
18	0.000250244140625\\
20	0.00028076171875\\
22	0.000274658203125\\
24	0.000244140625\\
26	0.0002197265625\\
28	0.000244140625\\
};
\addlegendentry{QWF};

\addplot [color=mycolor5,solid,line width=2.0pt,mark=o,mark options={solid}]
  table[row sep=crcr]{%
-10	0.2246826171875\\
-8	0.174444580078125\\
-6	0.116552734375\\
-4	0.067437744140625\\
-2	0.030950927734375\\
0	0.0098388671875\\
2	0.001849365234375\\
4	0.000140380859375\\
6	6.103515625e-06\\
8	0\\
10	0\\
12	0\\
14	0\\
16	0\\
18	0\\
20	0\\
22	0\\
24	0\\
26	0\\
28	0\\
};
\addlegendentry{CE [32]};

\addplot [color=mycolor4,dashed,line width=2.0pt,mark=o,mark options={solid}]
  table[row sep=crcr]{%
-10	0.2181640625\\
-8	0.16773681640625\\
-6	0.114373779296875\\
-4	0.07003173828125\\
-2	0.035870361328125\\
0	0.01444091796875\\
2	0.004522705078125\\
4	0.001287841796875\\
6	0.000244140625\\
8	4.8828125e-05\\
10	2.44140625e-05\\
12	1.220703125e-05\\
14	0\\
16	6.103515625e-06\\
18	0\\
20	0\\
22	0\\
24	0\\
26	0\\
28	0\\
};
\addlegendentry{WF-CE};

\addplot [color=mycolor6,dashed,line width=2.0pt]
  table[row sep=crcr]{%
-10	0.188372802734375\\
-8	0.1370849609375\\
-6	0.083636474609375\\
-4	0.042425537109375\\
-2	0.016046142578125\\
0	0.003533935546875\\
2	0.000433349609375\\
4	1.220703125e-05\\
6	0\\
8	0\\
10	0\\
12	0\\
14	0\\
16	0\\
18	0\\
20	0\\
22	0\\
24	0\\
26	0\\
28	0\\
};
\addlegendentry{WF, ideal};

\end{axis}
\end{tikzpicture}%
}
\caption{Uncoded BER performance for a MU-MIMO system with $N=64$ and $M=8$ with different precoding designs and QPSK signaling.}
\label{fig:ber_all_64_8_Q4}
\end{figure}
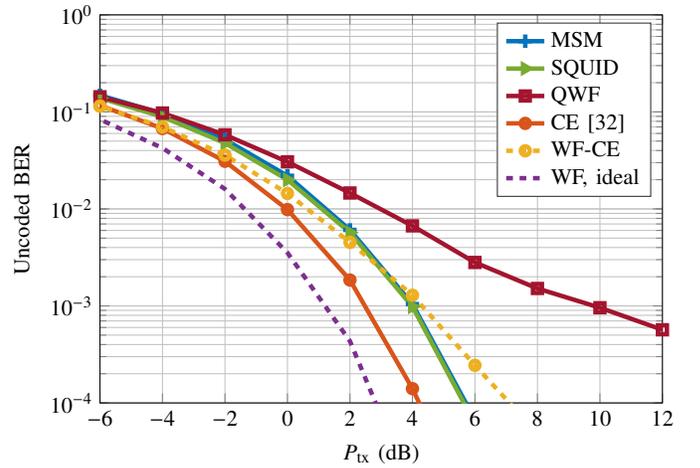

In the first simulation set, depicted in Fig.~\ref{fig:ber_all_64_8_Q4}, we assume full \gls{CSI}, choose QPSK modulation and compare the uncoded \gls{BER} as a function of the transmit power $P_{\text{tx}}$ for the following precoders
\begin{itemize}
\item the proposed \gls{MSM} method with $Q=4$,
\item the SQUID precoder presented in \cite{Jacobsson.2016} with $Q=4$,
\item the quantized \gls{WF} precoder denoted by "QWF" from \cite{Mezghani.2009} with $Q=4$,
\item the \gls{CE} precoder presented in \cite{Noll.2017} denoted by "CE \cite{Noll.2017}", with $Q=\infty$, where the precoding gain $\alpha$ is taken into account,
\item the \gls{WF} precoder followed by the \gls{CE} quantizer with $Q=\infty$ denoted by "WF-CE", and 
\item the \gls{WF} precoder in the ideal case denoted by "WF, ideal", where neither quantization nor the \gls{CE} constraint is applied to the transmit signal.
\end{itemize}
It can be seen that the \gls{CE} constraint leads to a loss, compared to the ideal case, of almost 2~dB at a \gls{BER} of $10^{-2}$ compared to WF and a loss of less than 1.5~dB when using the symbol-wise precoder proposed in \cite{Noll.2017}. The 1-bit quantization, which represents the \gls{QCE} case of $Q=4$, leads to more losses that depend on the precoder design. With the use of the linear precoder QWF a loss of more than 4~dB at a \gls{BER} of $10^{-2}$ is noticed. However, the non-linear precoders MSM and SQUID improve the performance drastically and show a loss of slightly more than $2~$dB compared to the ideal caseat the cost of higher computational
complexity. Nevertheless, the proposed \gls{MSM} method appears to be more efficient than SQUID as it is based on a pure linear programming formulation that has been intensively investigated in the literature.  

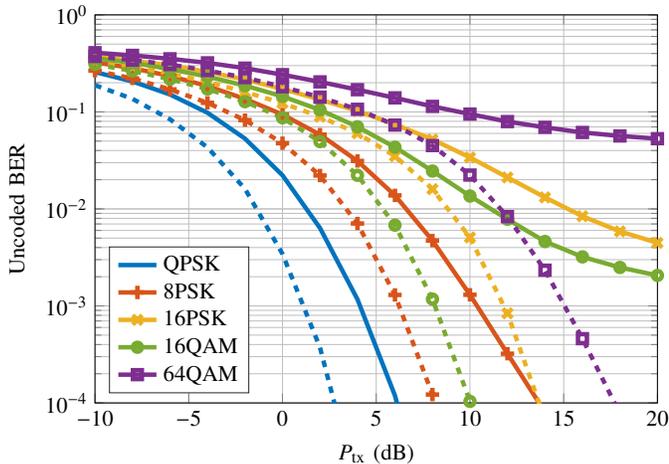
\begin{figure}
\centering
\resizebox{0.5\textwidth}{!} {
%
%
\definecolor{mycolor1}{rgb}{0.00000,0.45098,0.74118}%
\definecolor{mycolor2}{rgb}{0.85098,0.32941,0.10196}%
\definecolor{mycolor3}{rgb}{0.92941,0.69020,0.12941}%
\definecolor{mycolor4}{rgb}{0.47059,0.67059,0.18824}%
\definecolor{mycolor5}{rgb}{0.49020,0.18039,0.56078}%
\begin{tikzpicture}

\begin{axis}[%
width=\columnwidth,
height=0.25\textheight,
at={(0.758in,0.481in)},
scale only axis,
xmin=-10,
xmax=20,
xlabel={$P_{\text{tx}}$~(dB)},
xmajorgrids,
ymode=log,
ymin=1e-4,
ymax=1,
yminorticks=true,
ylabel={Uncoded BER},
ymajorgrids,
yminorgrids,
axis background/.style={fill=white},
legend style={at={(0.025,0.025)},anchor=south west,legend cell align=left,align=left,draw=white!15!black}
]

\addplot [color=mycolor1,dashed,line width=2.0pt,forget plot]
  table[row sep=crcr]{%
-10	0.188555908203125\\
-8	0.13634033203125\\
-6	0.083978271484375\\
-4	0.043048095703125\\
-2	0.016119384765625\\
0	0.003436279296875\\
2	0.000372314453125\\
4	1.220703125e-05\\
6	0\\
8	0\\
10	0\\
12	0\\
14	0\\
16	0\\
18	0\\
20	0\\
22	0\\
24	0\\
26	0\\
28	0\\
};

\addplot [color=mycolor1,solid,line width=2.0pt]
  table[row sep=crcr]{%
-10	0.257612915039062\\
-8	0.205874633789063\\
-6	0.151273193359375\\
-4	0.0980560302734375\\
-2	0.0530572509765625\\
0	0.0222235107421875\\
2	0.0063555908203125\\
4	0.0011676025390625\\
6	0.0001190185546875\\
8	4.2724609375e-06\\
10	0\\
12	0\\
14	0\\
};
\addlegendentry{QPSK};

\addplot [color=mycolor2,dashed,line width=2.0pt,mark=+,mark options={solid},mark size=3pt,forget plot]
  table[row sep=crcr]{%
-10	0.266499837239583\\
-8	0.221341959635417\\
-6	0.170662434895833\\
-4	0.124039713541667\\
-2	0.08231201171875\\
0	0.04755859375\\
2	0.0221110026041667\\
4	0.00707600911458333\\
6	0.00129801432291667\\
8	0.0001220703125\\
10	0\\
12	0\\
14	0\\
16	0\\
18	0\\
20	0\\
22	0\\
24	0\\
26	0\\
28	0\\
};

\addplot [color=mycolor2,solid,line width=2.0pt,mark=+,mark options={solid},mark size=3pt]
  table[row sep=crcr]{%
-10	0.325050862630208\\
-8	0.282132975260417\\
-6	0.2345166015625\\
-4	0.184654947916667\\
-2	0.13620361328125\\
0	0.0939408365885417\\
2	0.05833251953125\\
4	0.031212158203125\\
6	0.0137841796875\\
8	0.004718017578125\\
10	0.00130289713541667\\
12	0.000321858723958333\\
14	8.056640625e-05\\
};
\addlegendentry{8PSK};

\addplot [color=mycolor3,dashed,line width=2.0pt,mark=x,mark options={solid},mark size=3pt, forget plot]
  table[row sep=crcr]{%
-10	0.328897094726562\\
-8	0.29039306640625\\
-6	0.247500610351563\\
-4	0.20443115234375\\
-2	0.160501098632813\\
0	0.12293701171875\\
2	0.0894622802734375\\
4	0.0601898193359375\\
6	0.034954833984375\\
8	0.0159881591796875\\
10	0.0050811767578125\\
12	0.0008392333984375\\
14	6.7138671875e-05\\
16	0\\
18	0\\
20	0\\
22	0\\
24	0\\
26	0\\
28	0\\
};
\addplot [color=mycolor3,solid,line width=2.0pt,mark=x,mark options={solid},mark size=3pt]
  table[row sep=crcr]{%
-10	0.370986633300781\\
-8	0.338522644042969\\
-6	0.301407470703125\\
-4	0.258978881835937\\
-2	0.21545654296875\\
0	0.172501220703125\\
2	0.134166259765625\\
4	0.102020568847656\\
6	0.0744070434570313\\
8	0.0520046997070312\\
10	0.0339007568359375\\
12	0.0211187744140625\\
14	0.01314697265625\\
16	0.00840728759765625\\
18	0.00586090087890625\\
20	0.00446868896484375\\
22	0.00360107421875\\
24	0.0032098388671875\\
26	0.00292572021484375\\
28	0.002734375\\
};
\addlegendentry{16PSK};

\addplot [color=mycolor4,dashed,line width=2.0pt,mark=o,mark options={solid},forget plot]
  table[row sep=crcr]{%
-10	0.309283447265625\\
-8	0.264300537109375\\
-6	0.217343139648437\\
-4	0.172268676757812\\
-2	0.127468872070312\\
0	0.086944580078125\\
2	0.0494354248046875\\
4	0.0222137451171875\\
6	0.006805419921875\\
8	0.0011749267578125\\
10	0.000103759765625\\
12	0\\
14	0\\
16	0\\
18	0\\
20	0\\
22	0\\
24	0\\
26	0\\
28	0\\
};
\addplot [color=mycolor4,solid,line width=2.0pt,mark=o,mark options={solid}]
  table[row sep=crcr]{%
-10	0.357061767578125\\
-8	0.3206298828125\\
-6	0.276248168945313\\
-4	0.231195068359375\\
-2	0.186663818359375\\
0	0.14461669921875\\
2	0.104898071289063\\
4	0.0701690673828125\\
6	0.0432220458984375\\
8	0.024578857421875\\
10	0.013580322265625\\
12	0.007818603515625\\
14	0.0046142578125\\
16	0.003204345703125\\
18	0.0024932861328125\\
20	0.00206298828125\\
22	0.00184326171875\\
24	0.001605224609375\\
26	0.0015869140625\\
28	0.0014923095703125\\
};
\addlegendentry{16QAM};

\addplot [color=mycolor5,dashed,line width=2.0pt,mark=square,mark options={solid},forget plot]
  table[row sep=crcr]{%
-10	0.373470052083333\\
-8	0.342213948567708\\
-6	0.30787353515625\\
-4	0.267097981770833\\
-2	0.225687662760417\\
0	0.181683349609375\\
2	0.141874186197917\\
4	0.106368001302083\\
6	0.07315673828125\\
8	0.0449910481770833\\
10	0.0223347981770833\\
12	0.00832926432291667\\
14	0.00233561197916667\\
16	0.000459798177083333\\
18	7.73111979166667e-05\\
20	1.42415364583333e-05\\
22	0\\
24	2.03450520833333e-06\\
26	0\\
28	0\\
};

\addplot [color=mycolor5,solid,line width=2.0pt,mark=square,mark options={solid}]
table[row sep=crcr]{%
-10	0.407493082682292\\
-8	0.382661946614583\\
-6	0.3525390625\\
-4	0.320406087239583\\
-2	0.282548014322917\\
0	0.241949462890625\\
2	0.203940836588542\\
4	0.169429524739583\\
6	0.139601643880208\\
8	0.113948567708333\\
10	0.0949747721354167\\
12	0.079437255859375\\
14	0.0693277994791667\\
16	0.06136474609375\\
18	0.0565104166666667\\
20	0.0529500325520833\\
22	0.051422119140625\\
24	0.0499471028645833\\
26	0.0494059244791667\\
28	0.0488321940104167\\
};
\addlegendentry{64QAM};

\end{axis}
\end{tikzpicture}%
}
\caption{Uncoded BER performance for a MU-MIMO system with $N=64$ and $M=8$ for different modulation schemes and $Q=4$: the dashed lines represent the uncoded BER results obtained in the case of WF, ideal.}
\label{fig:ber_64_8_all_Q4}
\end{figure}

\begin{figure}
\centering
\resizebox{0.5\textwidth}{!} {
%
%
\definecolor{mycolor1}{rgb}{0.00000,0.45098,0.74118}%
\definecolor{mycolor2}{rgb}{0.85098,0.32941,0.10196}%
\definecolor{mycolor3}{rgb}{0.47059,0.67059,0.18824}%
\definecolor{mycolor4}{rgb}{0.92941,0.69020,0.12941}%
\definecolor{mycolor5}{rgb}{0.49020,0.18039,0.56078}%
\begin{tikzpicture}

\begin{axis}[%
width=\columnwidth,
height=0.25\textheight,
at={(0.758in,0.481in)},
scale only axis,
xmin=-10,
xmax=20,
xlabel={$P_{\text{tx}}$~(dB)},
xmajorgrids,
ymode=log,
ymin=1e-4,
ymax=1,
yminorticks=true,
ylabel={Uncoded BER},
ymajorgrids,
yminorgrids,
axis background/.style={fill=white},
legend style={at={(0.025,0.025)},anchor=south west,legend cell align=left,align=left,draw=white!15!black}
]
\addplot [color=mycolor1,dashed,line width=2.0pt,forget plot]
  table[row sep=crcr]{%
-10	0.188555908203125\\
-8	0.13634033203125\\
-6	0.083978271484375\\
-4	0.043048095703125\\
-2	0.016119384765625\\
0	0.003436279296875\\
2	0.000372314453125\\
4	1.220703125e-05\\
6	0\\
8	0\\
10	0\\
12	0\\
14	0\\
16	0\\
18	0\\
20	0\\
22	0\\
24	0\\
26	0\\
28	0\\
};
\addplot [color=mycolor1,solid,line width=2.0pt]
  table[row sep=crcr]{%
-10	0.237472534179688\\
-8	0.18427978515625\\
-6	0.129391479492187\\
-4	0.0782965087890625\\
-2	0.0377301025390625\\
0	0.012957763671875\\
2	0.00282958984375\\
4	0.000316162109375\\
6	1.46484375e-05\\
8	0\\
10	0\\
12	0\\
14	0\\
};
\addlegendentry{QPSK};

\addplot [color=mycolor2,dashed,line width=2.0pt,mark=+,mark options={solid},mark size=3pt,forget plot]
  table[row sep=crcr]{%
-10	0.266499837239583\\
-8	0.221341959635417\\
-6	0.170662434895833\\
-4	0.124039713541667\\
-2	0.08231201171875\\
0	0.04755859375\\
2	0.0221110026041667\\
4	0.00707600911458333\\
6	0.00129801432291667\\
8	0.0001220703125\\
10	0\\
12	0\\
14	0\\
16	0\\
18	0\\
20	0\\
22	0\\
24	0\\
26	0\\
28	0\\
};
\addplot [color=mycolor2,solid,line width=2.0pt,mark=+,mark options={solid},mark size=3pt]
  table[row sep=crcr]{%
-10	0.308694254557292\\
-8	0.263629150390625\\
-6	0.214412841796875\\
-4	0.164312337239583\\
-2	0.117008870442708\\
0	0.0765262858072917\\
2	0.043284912109375\\
4	0.0195149739583333\\
6	0.006407470703125\\
8	0.00131673177083333\\
10	0.000150146484375\\
12	6.91731770833333e-06\\
14	8.13802083333333e-07\\
};
\addlegendentry{8PSK};
\addplot [color=mycolor4,dashed,line width=2.0pt,mark=x,mark options={solid},mark size=3pt,forget plot]
  table[row sep=crcr]{%
-10	0.328897094726562\\
-8	0.29039306640625\\
-6	0.247500610351563\\
-4	0.20443115234375\\
-2	0.160501098632813\\
0	0.12293701171875\\
2	0.0894622802734375\\
4	0.0601898193359375\\
6	0.034954833984375\\
8	0.0159881591796875\\
10	0.0050811767578125\\
12	0.0008392333984375\\
14	6.7138671875e-05\\
16	0\\
18	0\\
20	0\\
22	0\\
24	0\\
26	0\\
28	0\\
};
\addplot [color=mycolor4,solid,line width=2.0pt,mark=x,mark options={solid},mark size=3pt]
  table[row sep=crcr]{%
-10	0.359107971191406\\
-8	0.324346923828125\\
-6	0.284414672851563\\
-4	0.240962219238281\\
-2	0.196108703613281\\
0	0.153583068847656\\
2	0.116614379882812\\
4	0.0845578002929687\\
6	0.0570639038085937\\
8	0.03376708984375\\
10	0.01689453125\\
12	0.00678497314453125\\
14	0.00223541259765625\\
16	0.0006219482421875\\
18	0.00017791748046875\\
20	5.79833984375e-05\\
22	2.50244140625e-05\\
24	1.556396484375e-05\\
26	1.190185546875e-05\\
28	8.23974609375e-06\\
};
\addlegendentry{16PSK};
\addplot [color=mycolor3,dashed,line width=2.0pt,mark=o,mark options={solid},forget plot]
  table[row sep=crcr]{%
-10	0.309283447265625\\
-8	0.264300537109375\\
-6	0.217343139648437\\
-4	0.172268676757812\\
-2	0.127468872070312\\
0	0.086944580078125\\
2	0.0494354248046875\\
4	0.0222137451171875\\
6	0.006805419921875\\
8	0.0011749267578125\\
10	0.000103759765625\\
12	0\\
14	0\\
16	0\\
18	0\\
20	0\\
22	0\\
24	0\\
26	0\\
28	0\\
};
\addplot [color=mycolor3,solid,line width=2.0pt,mark=o,mark options={solid}]
  table[row sep=crcr]{%
-10	0.345565795898438\\
-8	0.303884887695313\\
-6	0.258770751953125\\
-4	0.21329345703125\\
-2	0.16732177734375\\
0	0.125442504882813\\
2	0.0864288330078125\\
4	0.05302734375\\
6	0.0288543701171875\\
8	0.0135467529296875\\
10	0.0061676025390625\\
12	0.002972412109375\\
14	0.0015106201171875\\
16	0.0008209228515625\\
18	0.0005889892578125\\
20	0.0003753662109375\\
22	0.0003509521484375\\
24	0.0003326416015625\\
26	0.0002838134765625\\
28	0.0002532958984375\\
};
\addlegendentry{16QAM};

\addplot [color=mycolor5,dashed,line width=2.0pt,mark=square,mark options={solid},forget plot]
  table[row sep=crcr]{%
-10	0.373470052083333\\
-8	0.342213948567708\\
-6	0.30787353515625\\
-4	0.267097981770833\\
-2	0.225687662760417\\
0	0.181683349609375\\
2	0.141874186197917\\
4	0.106368001302083\\
6	0.07315673828125\\
8	0.0449910481770833\\
10	0.0223347981770833\\
12	0.00832926432291667\\
14	0.00233561197916667\\
16	0.000459798177083333\\
18	7.73111979166667e-05\\
20	1.42415364583333e-05\\
22	0\\
24	2.03450520833333e-06\\
26	0\\
28	0\\
};
\addplot [color=mycolor5,solid,line width=2.0pt,mark=square,mark options={solid}]
  table[row sep=crcr]{%
-10	0.397859700520833\\
-8	0.370003255208333\\
-6	0.339495849609375\\
-4	0.305037434895833\\
-2	0.265061442057292\\
0	0.224471028645833\\
2	0.185135904947917\\
4	0.150030517578125\\
6	0.120501708984375\\
8	0.0948506673177083\\
10	0.075653076171875\\
12	0.060186767578125\\
14	0.0496134440104167\\
16	0.0430765787760417\\
18	0.0390909830729167\\
20	0.0368855794270833\\
22	0.0355204264322917\\
24	0.034661865234375\\
26	0.0338887532552083\\
28	0.0336222330729167\\
};
\addlegendentry{64QAM};

\end{axis}
\end{tikzpicture}%
}
\caption{Uncoded BER performance for a MU-MIMO system with $N=64$ and $M=8$ for different modulation schemes and $Q=8$: the dashed lines represent the uncoded BER results obtained in the case of WF, ideal.}
\label{fig:ber_64_8_all_Q8}
\end{figure}
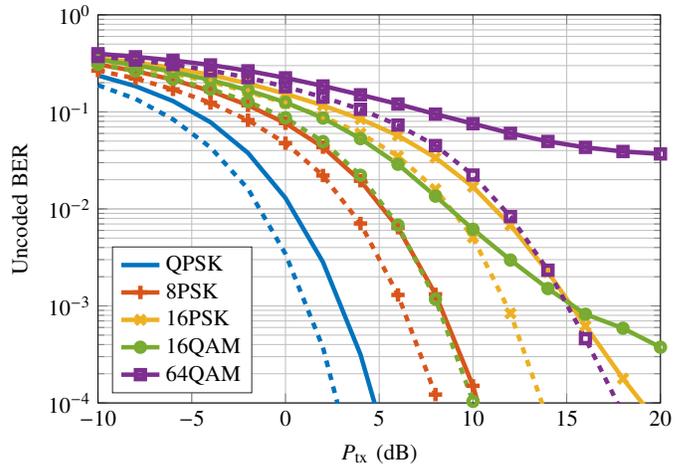

In the second simulation set, depicted in Fig.~\ref{fig:ber_64_8_all_Q4} and Fig.~\ref{fig:ber_64_8_all_Q8}, the uncoded BER is plotted as a function of the transmit power $P_{\text{tx}}$ using the \gls{MSM} precoder for different modulation schemes and two different values of $Q$: $Q=4$ and $Q=8$. Higher values of $Q$ are omitted since the obtained results do not differ much from the case of $Q=8$. In addition, it is beneficial in terms of computational complexity and power consumption to keep $Q$ as small as possible.
As expected, the higher the number of symbols in the modulation scheme, the higher the \gls{BER} for a given $P_{\text{tx}}$ value. However, the increase of the \gls{DAC} resolution $q$ and thus the increase of $Q$ leads to a performance improvement, which depends on the modulation scheme. Interestingly, the 16QAM results outperform the 16PSK results with a gain of almost 4~dB at a \gls{BER} of $10^{-2}$ for the case of $Q=4$, whereas in the case of $Q=8$ the gain reduces to 3~dB and the 16PSK modulation outperforms the 16QAM   for transmit power values larger than 15~dB.

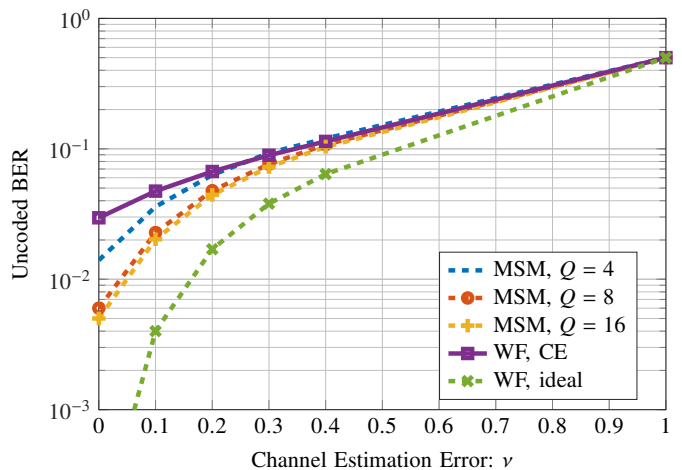
\begin{figure}
\centering
\resizebox{0.5\textwidth}{!} {
\definecolor{mycolor1}{rgb}{0.00000,0.44700,0.74100}%
\definecolor{mycolor2}{rgb}{0.85000,0.32500,0.09800}%
\definecolor{mycolor3}{rgb}{0.92900,0.69400,0.12500}%
\definecolor{mycolor4}{rgb}{0.49400,0.18400,0.55600}%
\definecolor{mycolor5}{rgb}{0.47059,0.67059,0.18824}%

\begin{tikzpicture}

\begin{axis}[%
width=\columnwidth,
height=0.25\textheight,
at={(0.758in,0.481in)},
scale only axis,
xmin=0,
xmax=1,
xlabel={Channel Estimation Error: $\nu$},
xmajorgrids,
every outer y axis line/.append style={white!15!black},
every y tick label/.append style={font=\color{white!15!black}},
ymode=log,
ymin=1e-03,
ymax=1,
yminorticks=true,
ylabel={Uncoded BER},
ymajorgrids,
yminorgrids,
title style={font=\bfseries},
legend style={at={(0.6,0.025)},anchor=south west,legend cell align=left,align=left,draw=white!15!black}
]
\addplot [color=mycolor1,dashed,line width=2.0pt]
table[row sep=crcr]{%
0	0.014\\
0.1	0.036\\
0.2	0.063\\
0.3	0.093\\
0.4	0.12\\
1	0.5\\
};
\addlegendentry{MSM, $ Q=4$};
\addplot [color=mycolor2,dashed,line width=2.0pt,mark=o,mark options={solid}]
table[row sep=crcr]{%
0	0.006\\
0.1	0.0228\\
0.2	0.04775\\
0.3	0.07635\\
0.4	0.107642\\
1	0.5\\
};
\addlegendentry{MSM, $ Q=8$};

\addplot [color=mycolor3,dashed,line width=2.0pt,mark=+,mark options={solid},mark size=3pt]
table[row sep=crcr]{%
0	0.005\\
0.1	0.020258\\
0.2	0.04437\\
0.3	0.07253\\
0.4	0.103505\\
1	0.5\\
};
\addlegendentry{MSM, $ Q=16$};

\addplot [color=mycolor4,solid,line width=2.0pt,mark=square,mark options={solid}]
table[row sep=crcr]{%
0	0.0296\\
0.1	0.0474\\
0.2	0.067\\
0.3	0.0893\\
0.4	0.114\\
1	0.5\\
};
\addlegendentry{WF, CE};

\addplot [color=mycolor5,dashed,line width=2.0pt,mark=x,mark options={solid},mark size=3pt]
table[row sep=crcr]{%
0	0.0001\\
0.1	0.004\\
0.2	0.017\\
0.3	0.038\\
0.4	0.064\\
1	0.5\\
};
\addlegendentry{WF, ideal};

\end{axis}
\end{tikzpicture}%
}
\caption{Uncoded BER performance as a function of the channel estimation error variance for 16\gls{QAM} signaling and $P_{\text{tx}}=10$~dB.}
\label{fig:ber_vs_csi}
\end{figure}

The third simulation set, depicted in Fig.~\ref{fig:ber_vs_csi}, addresses the system performance in the presence of channel estimation errors. The estimated channel is defined as
\begin{align}
\vH_{\nu} = \sqrt{1-\nu} \vH + \sqrt{\nu} \vGamma,
\end{align}
where $\vGamma$ is a random matrix with \gls{iid} zero-mean and unit-variance entries.
We can see that the performance of the proposed \gls{MSM} precoder in the case of erroneous channel estimation is still better than the linear \gls{WF} followed by the \gls{CE} quantizer with $Q = \infty$. 

In the last simulation set, we counted the average number of iterations required by the \gls{MSM} precoder. The results are summarized in Table \ref{tab:Nbiter}, where we observe that around 50 iterations are required for all different modulation schemes for $Q=4$ and more than 100 iterations for $Q>4$.
\begin{table}[h!]
  \begin{center}
    \caption{Average number of iterations of the MSM precoder.}
    \label{tab:Nbiter}
    \begin{tabular}{|c|c|c|c|} 
 	  \hline
       Nb. of iter & $Q=4$ & $Q=8$ & $Q=16$\\
      \hline 
      QPSK  & 45.77 & 121.05 & 187.63\\
      8PSK  & 50.15 & 123.91 & 191.55\\
      16PSK & 54.94 & 128.74  & 199.61\\
      16QAM &43.25 & 120.42& 187.32\\
      64QAM &43.04 &120.30&188.30\\
      \hline
    \end{tabular}
  \end{center}
\end{table}

\section{Conclusion}
\label{sec:conclusion}
We proposed a symbol-wise precoder for a massive \gls{MU}-\gls{MIMO} downlink system with coarsely \gls{QCE} signals at the transmit antennas. The \gls{CE} constraint is motivated by the high \gls{PA} power efficiency for \gls{CE} input signals, and the coarse quantization provides further power savings due to the use of the low-resolution \glspl{DAC}. The \gls{MSM} precoder is based on maximizing the safety margin to the receiver decision thresholds taking the \gls{QCE} into account. When relaxing the \gls{QCE} constraint to a convex set, the optimization problem can be formulated as a linear programming problem, and thus can be efficiently solved via a number of methods. The proposed precoding method comprises both \gls{PSK} and \gls{QAM} modulation schemes.

\bibliographystyle{IEEEtran}
\bibliography{main}

\end{document}